\documentclass{article}

\usepackage[english]{babel}

\usepackage[letterpaper,top=2cm,bottom=2cm,left=3cm,right=3cm,marginparwidth=1.75cm]{geometry}
\usepackage{amsmath}
\newenvironment{tablenotes}{
    \begin{flushleft} \small
    \textit{Note:} 
}{
    \end{flushleft}
}
\usepackage{setspace}
\usepackage{placeins}
\usepackage{enumitem}
\usepackage{natbib}
\usepackage{amssymb}
\usepackage{gensymb}
\usepackage{siunitx}
\PassOptionsToPackage{hyphens}{url}
\usepackage[colorlinks=true, linkcolor=black, citecolor=black, urlcolor=black]{hyperref}
\usepackage[utf8]{inputenc}
\usepackage{csquotes}
\usepackage{booktabs}
\usepackage{longtable}
\usepackage{adjustbox}
\usepackage{array}
\usepackage{url}
\usepackage[font=normalsize]{caption}
\usepackage{subcaption}
\usepackage{arydshln}
\renewcommand{\topfraction}{0.9}	
	
\renewcommand{\floatpagefraction}{0.7}	% require fuller float pages
\newcommand{\bitem}[1]{\item \textbf{#1}}

\begin{document}

\title{Shifting Work Patterns with Generative AI\thanks{Corresponding author: eldillon@microsoft.com. $^\dagger$ denotes equal contribution. We thank the Microsoft Customer Research Program, especially Alexia Cambon, Sida Peng, Neha Shah, Chris Gideon, Modern Work Marketing, the Office of Applied Research, and company partners for help carrying out this experiment, Abigail Atchison, Roman Basko, and Fabio Vera for superb data science support, Jack Cenatempo, Esther Plotnick, and Will Wang for excellent research assistance. We thank 3 anonymous referees, Tatyana Deryugina, Seema Jayachandran, Rem Koning, Anders Humlum, and Danielle Li for helpful comments.}}

\author{\\ %
\begin{tabular}{cc}
Eleanor Wiske Dillon$^\dagger$ & Sonia Jaffe$^\dagger$\tabularnewline
{\small{}Microsoft Research} & {\small{}Microsoft Research}\tabularnewline
\bigskip \tabularnewline
Nicole Immorlica & Christopher T. Stanton\tabularnewline
{\small{}Microsoft Research} & {\small{}Harvard Business School,
} \tabularnewline & {\small{}
NBER, and CEPR}\tabularnewline
\end{tabular} 
\bigskip
}
\date{October 28, 2025}
\date{\today}

\maketitle

\begin{abstract}

We present evidence from a field experiment across 66 firms and 7,137 knowledge workers. Workers were randomly selected to access a generative AI tool integrated into applications they already used at work for email, meetings, and writing. In the second half of the 6-month experiment, the 80\% of treated workers who used this tool spent two fewer hours on email each week and reduced their time working outside of regular hours. Apart from these individual time savings, we do not detect shifts in the quantity or composition of workers' tasks resulting from individual-level AI provision.

\end{abstract}

\onehalfspacing

Generative AI has opened new possibilities for technology to assist with or automate a variety of tasks.  Early studies have already shown that generative AI increases worker productivity in targeted tasks \citep[e.g.][]{brynjolfsson2023generative, peng2023impact,noy2023experimental, dell2023navigating}, and advances in this technology have the potential to shift the nature of jobs \citep{eloundou2024gpts, tomlinson2025working} and perhaps the macroeconomy \citep{autor2024applying,acemoglu2025simple}. In lab experiments and early field studies on generative AI, participants often focused on completing a single task with targeted, well-suited tools. In everyday settings, workers must first learn which parts of their job benefit from the use of new tools and then change their work habits. This dynamic helps to explain why the transition from the creation of new technologies to effective, widespread adoption in the workplace often takes time \citep{brynjolfsson2021productivity}.

In this paper, we present field evidence on the adoption of generative AI tools and their effects on the way knowledge workers do their jobs. Between September 2023 and October 2024, Microsoft partnered with 66 large firms to run a cross-industry field experiment to measure how access to an integrated generative AI tool, Microsoft 365 Copilot (hereafter, Copilot), changed work patterns. Each firm recruited at least 100 workers for random assignment to treatment and control groups. Treated workers received access to this new generative AI tool while controls continued working with their current technologies. Participants in the study came from varied occupations, but all did work that used the Microsoft Office tools into which Copilot was integrated (e.g., emailing, video meetings, chat, document creation). Firms were asked to maintain the random assignment for six months, giving treated workers time to explore the new tool and integrate it into their daily work.

The study took place during the pilot rollout of Copilot, when firms could purchase licenses for only a small fraction of their workers.  
Treated workers were free to use or not use the tool during their regular work. Although treated and control workers may have used other generative AI tools (for example, ChatGPT had been  available for nearly a year by the beginning of the study), Copilot integrated AI into applications the workers already used regularly in their jobs. This integrated tool allowed new uses, like summarizing email inboxes or searching for local information, and likely reduced the effort to deploy generative AI for writing documents and emails.

We use telemetry data collected by Microsoft products to track workers' behavior during the experiment and for several months beforehand. Our primary outcome measures are related to time use (e.g. time spent in meetings or reading emails) and quantities of activities (e.g., emails read, meetings attended, documents authored) within the Microsoft Office suite, as we do not observe the content of any work nor any measures of productivity or performance evaluation.

Over 90\% of the workers assigned to receive Copilot used it at least once in the 6 month period after treatment. Usage peaked within the first 5 weeks of treatment assignment, possibly because of training, novelty, or trial motives. To provide evidence on the effects of AI use when workers were already familiar with the tool, our main analysis focuses on outcomes at least 12 weeks after treatment assignment, where weekly average use stabilized at just under 40\% of treated workers.

Workers varied in the intensity of their Copilot use. Firm-level differences explain the largest share of this variation, followed by workers' pre-period behavior. This variation is consistent with other surveys of AI adoption across firms \citep{bonney2024tracking, mcelheran2024ai}, but it is notable in our more homogenous sample of large firms that were interested in being early adopters of this tool and participating in a study of its use. Firm characteristics like industry capture only some of the firm effects, suggesting an important role for firm-specific managerial practices or training that we cannot measure \citep{mcelheran2025rise}. 

Workers who used Copilot exhibited some meaningful changes in work behavior. Intention-to-treat estimates suggest workers with Copilot reduced their time using Outlook for email by 1.4 hours per week, a 12\% reduction relative to the pre-period mean. The Local Average Treatment Effect (LATE) estimate from instrumenting Copilot use with treatment assignment implies a 2 hour (17\%) reduction in Outlook time. Copilot users also consolidated when they worked on email, opening up 1.5 hours each week of extended work time without Outlook activity.\footnote{Appendix \ref{app:outcomes} details how we define each of these outcomes.}  In contrast, we find no significant average changes in total time spent in video meetings or writing documents.

We see clear evidence that workers did some of their existing work tasks more quickly, but do not detect meaningful shifts in workers' tasks \citep{autor2024applying} or innovations that reorganize work \citep{brynjolfsson2000beyond}. In particular, treated and control workers replied to the same number of email threads, participated in the same number of Teams meetings, and completed the same number of Word documents.\footnote{While not reported, we also find little change in activity in PowerPoint or Excel.} While we cannot observe a complete accounting of activities during work, given the occupations of the studied workers and their employment at Office-heavy firms, it would be surprising if they took on major new work responsibilities while leaving no trace of changing behavior in these digital applications. Treated workers did reduce their time spent in digital applications outside of their regular working hours by about 0.25 hours-per-week (9\%), suggesting that some of the time saved on email returned to workers. We do not find any negative spillover effects on the work patterns of treated workers' close coworkers, suggesting that the time savings we document do not come at the expense of work quality.

This study highlights an under-appreciated role for generative AI in real world production processes. In our sample, workers were most likely to access Copilot through Teams or Outlook. The Teams use in particular demonstrates the value of generative AI tools for summarizing and organizing dispersed information. While this role of AI has received less attention than text or code generation, its importance is supported by other recent studies. For example, \citet{tomlinson2025working} identify information gathering as the most common use of a generative AI chat tool. \citet{blandin2024rapid} find that managers, who spend more time in meetings than other workers \citep{impink2025communication}, are tied with computer/mathematical occupations as the groups most likely to use generative AI at work.

Our work adds to studies on the effects of generative AI in real workplaces \citep{brynjolfsson2023generative,cui2024impact,otis2023uneven}, while complementing studies that use controlled tasks in a lab setting.\footnote{See, for example, \cite{peng2023impact}, \cite{Spatharioti2024search}, \cite{vaithilingam2022expectation}, \cite{campero2022}, \cite{noy2023experimental}, \cite{dell2023navigating}, \cite{cambon2023early}.} We also contribute to the broader literature on technology diffusion \citep{comin2010exploration,forman2018technology}. Surveys indicate rapid adoption of generative AI at work across a broader set of workers than those considered in lab or early field settings. \cite{humlum2025unequal} find 41\% of Danish workers in exposed occupations used ChatGPT at work by the end of 2023 and \cite{blandin2024rapid} find 24\% of all U.S. workers used generative AI at work by August 2024. In follow-up work, \cite{humlum2025large} find that AI chatbots at work have not had measurable impacts on worker earnings or hours  worked, largely consistent with our mixed results on worker behavior. This study bridges some of the gaps between these two lines of research: measuring both take up and impacts of use for a generative AI tool on a broad set of workers, albeit at the cost of clear measures of worker productivity.

\section{Treatment Assignment and Patterns of Copilot Adoption}

\subsection{Experimental Design}

In fall 2023, Microsoft offered select large firms the option to purchase exactly 300 Copilot licenses. Copilot was not available to smaller firms or to individual consumers, and firms could not purchase additional licenses beyond 300. Within this group, Microsoft's marketing team recruited firms to join the study with the understanding that they would allocate at least 50 of these 300 licenses to the experiment and recruit at least twice as many participating workers. The Copilot licenses were assigned at random among experiment-eligible workers, but treated workers were not required to use the tools or deploy them for specific work tasks. In exchange for participating in the study, firms received early access to findings from the experiment, including customized firm-level reports. They did not receive discounts or access to additional licenses. 

Appendix \ref{app:exp} includes more details on the 7,137 participating workers (3,684 treated) across 66 firms. The firms span a range of industries (see Figure \ref{fig:firm_ind}). All are very large and most operate in multiple countries. Firms opted into the study in response to only modest incentives and may be more proactive in adopting new technologies than the average firm. 

Firms varied in how they recruited participating workers, with some targeting specific functions or workers with a track record of adopting new technologies early and others creating an open call for participation.\footnote{We suspect that our sample excludes the highest-priority Copilot users, who would have received licenses with certainty rather than by lottery. However, our estimates may still overstate population average treatment effects if firms targeted the experiment to workers where they expected to see large gains.} Appendix Table \ref{table:occupation_table} tabulates occupations for the subset of workers for whom we have that information. As expected, the sample is dominated by knowledge workers who regularly do the kinds of emailing and word processing tasks for which Copilot was designed. Appendix Table \ref{table:balance_table} summarizes pre-period work behaviors for treated and control workers in the sample. Unfortunately, our data use agreements do not allow us to compare these sampled workers with the broader set of workers at the firms, but we can compare our sample with the knowledge workers studied in \citet{impink2025communication}.  They document workers' email and meeting behavior in a sample of 102 large firms around 2019. Our sample of workers seems roughly comparable to theirs. For example, their studied workers attended a mean of 39 meetings per month while ours attended 43. Nonetheless, we caveat that participating workers may not be representative of the overall worker population at participating firms, or of knowledge workers more broadly.

\subsection{Adoption}
\begin{figure}[htbp]
    \centering
     
    \begin{subfigure}{0.46\textwidth}
        \includegraphics[width=\textwidth]{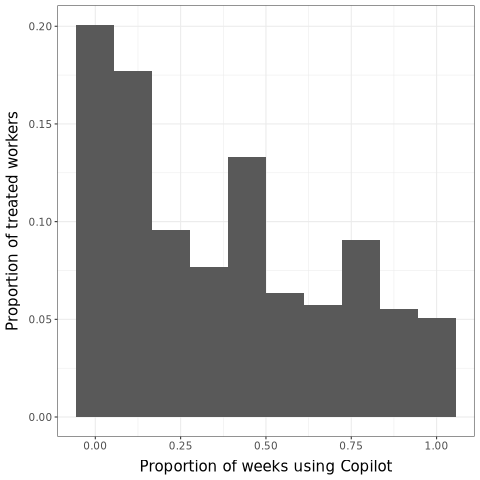}
        \caption{Usage Rate by Individual%rate by individual
        \label{fig:individual_usage}}
    \end{subfigure}
    \hfill
        \begin{subfigure}{0.46\textwidth}   
        \includegraphics[width=\textwidth]{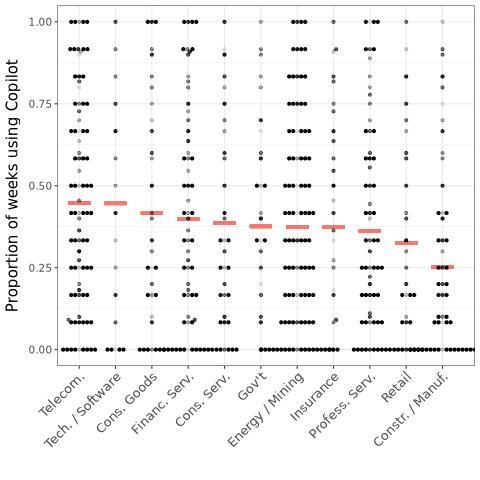}
        \caption{Usage Rate by Industry %rate by industry %- Across Firms
        \label{fig:industry_usage}}
    \end{subfigure}
        \medskip
    \begin{subfigure}{0.46\textwidth}
        \includegraphics[width=\textwidth]{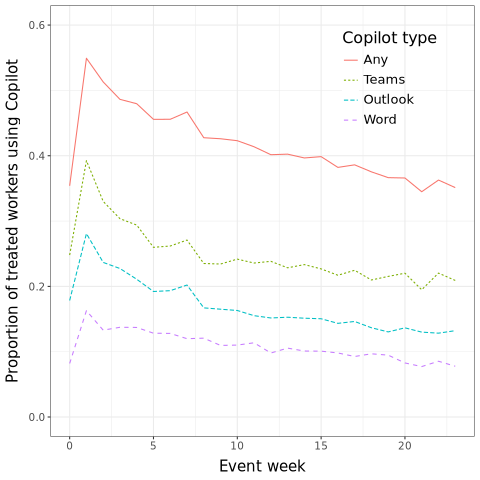}
        \caption{Usage Rate by App%rate by app %- Across Firms
        \label{fig:app_usage}}
    \end{subfigure}
\hfill
    \begin{subfigure}{0.46\textwidth}
        \includegraphics[width=\textwidth]{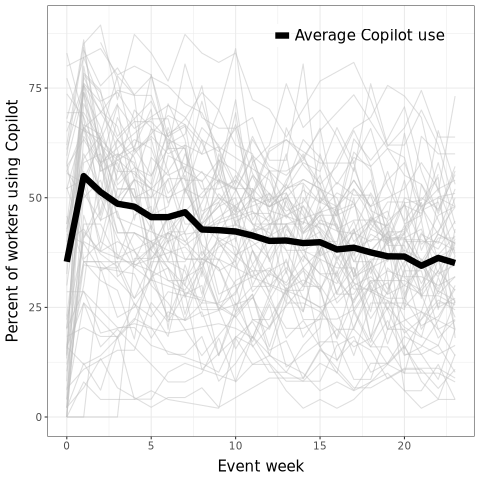}
        \caption{Usage Rate by Firm%rate by app %- Across Firms
        \label{fig:firm_usage}}
    \end{subfigure}

   \caption{Copilot Usage in the Treatment Group}
   \label{fig:usage_combined}
   \begin{tablenotes}
       Panels (a) and (b) show the proportion of weeks in post-period months 4-6 that each worker used Copilot overall (a) and by industry (b). In Panel (b), each solid dot represents 8 workers; lighter dots represent fewer than 8 workers. Mean industry usage rates are highlighted in red. Panel (c) shows the share of treated workers using Copilot in each post-period week, through any application and specifically through Teams, Outlook, and Word. Panel (d) the share of treated workers using Copilot in each post-period week separately for each participating firm. The black line in panel (d) shows the average across firms, replicating the red line in panel (c).
   \end{tablenotes}
\end{figure}

The median treated worker used Copilot in 39\% of the weeks they had access to it during the study (33\% of the weeks in months 4-6).\footnote{In our data, use means taking an active action to engage with a Copilot tool through any of the Copilot-enabled applications: Outlook, Teams, Word, Excel, PowerPoint, and the separate M365 Chat. Having Copilot open by default is not sufficient. We do not observe any information about prompts or the type of use, only which application-specific Copilot tools were engaged.}  Figure \ref{fig:individual_usage} shows the distribution of individual usage rates over months 4-6; 5\% of treated workers used Copilot every week, while 20\% never used it at all. As shown in Figure \ref{fig:app_usage}, overall usage peaked at 55\% of treated workers in the first weeks of the study as workers explored the new tool and, at some firms, participated in introductory training sessions.

Figure \ref{fig:app_usage} also shows Copilot use separately for Outlook, Teams, and Word. We focus on time use and behavior in these applications because the early phase of prompt-based generative AI was most applicable to tasks in these tools. Workers were most likely to use Copilot in Teams, which can analyze transcripts of meetings to generate summaries, list follow-up items, or answer user queries. The next most common use was through Outlook, where Copilot can summarize email chains, flag emails that require a response, answer questions more flexibly than a keyword search, and generate draft emails in response to prompts. The Word Copilot can draft text from prompts and answer questions about document content. The Word Copilot had the lowest usage of the three we study, partially because not all experimental workers used Word regularly in their jobs (see Appendix \ref{sec:dataapp}). Our regressions for Word-based outcomes limit the sample to the 2,525 study participants who used Word regularly in the pre-period (39\% of our primary analysis sample). Copilots for PowerPoint and Excel had lower usage rates at the time of the study. The application-specific Copilot usage rates displayed in the figure sum to more than the overall usage rate because, as illustrated in Appendix Table \ref{tab:use_correl}, most workers who used Copilot used it in multiple applications. For example, 80\% of workers who used Copilot in Outlook in months 4-6 also used Copilot in Teams at least once in the same period.

Firms are a key driver of workers' propensity to use Copilot. As shown in Figure \ref{fig:firm_usage}, usage rates varied widely and consistently across firms, with average usage rates among treated workers as low as 6.3\% per week at one firm over the final 3 months of the study and as high as 70\% at another. Some of this variation stems from differences across industries. In telecommunications and technology firms, 45\% of treated workers  used Copilot each week, but only 25\% did at construction and manufacturing firms (see Figure \ref{fig:industry_usage}). However, considerable variation in usage rates exists within industries.

\begin{table}[htbp]
    \centering
    \caption{Predicting Adoption \label{table:horserace}}
    \resizebox{\textwidth}{!}{%
\begingroup
\centering
\begin{tabular}{lccccccccc}
   \midrule \midrule
   Dependent Variable: & \multicolumn{9}{c}{Individual average Copilot use}\\
   Model:                                          & (1)     & (2)     & (3)     & (4)     & (5)   & (6)     & (7)   & (8)   & (9)\\  
   \midrule
   % \emph{Variables}\\
   Constant                                        & 0.282   & 0.279   & 0.142   & 0.116   &       &         &       &       &   \\   
                                                   & (0.011) & (0.012) & (0.020) & (0.022) &       &         &       &       &   \\   
   Emails read (thousands)                         & -0.207  & -0.203  &         & -0.183  &       & -0.198  &       &       & -0.191\\   
                                                   & (0.060) & (0.060) &         & (0.066) &       & (0.063) &       &       & (0.065)\\   
   Unique email threads replied to (thousands)     & -0.853  & -0.862  &         & -0.572  &       & -0.572  &       &       & -0.492\\   
                                                   & (0.559) & (0.559) &         & (0.588) &       & (0.567) &       &       & (0.578)\\   
   Teams meeting time (hours)                      & 0.012   & 0.012   &         & 0.011   &       & 0.011   &       &       & 0.011\\   
                                                   & (0.003) & (0.003) &         & (0.003) &       & (0.003) &       &       & (0.003)\\   
   Total Teams meetings attended                   & 0.007   & 0.007   &         & 0.002   &       & 0.002   &       &       & 0.002\\   
                                                   & (0.002) & (0.002) &         & (0.002) &       & (0.002) &       &       & (0.002)\\   
   Word total session time (hours)                 & 0.017   & 0.017   &         & 0.010   &       & 0.011   &       &       & 0.011\\   
                                                   & (0.005) & (0.005) &         & (0.005) &       & (0.005) &       &       & (0.005)\\   
   Others' documents read/edited before completion & 0.038   & 0.037   &         & 0.031   &       & 0.030   &       &       & 0.031\\   
                                                   & (0.013) & (0.013) &         & (0.013) &       & (0.013) &       &       & (0.013)\\   
   Number of email recipients (tens)               & -0.002  & -0.002  &         & -0.006  &       & -0.005  &       &       & -0.006\\   
                                                   & (0.004) & (0.004) &         & (0.004) &       & (0.004) &       &       & (0.004)\\   
   Number of unique people met 1-1 (total, tens)   & 0.003   & 0.003   &         & 0.003   &       & 0.003   &       &       & 0.002\\   
                                                   & (0.005) & (0.005) &         & (0.005) &       & (0.005) &       &       & (0.005)\\   
   Share of coworkers with Copilot license         &         & 0.013   &         & 0.054   &       & 0.051   &       &       & 0.051\\   
                                                   &         & (0.023) &         & (0.023) &       & (0.024) &       &       & (0.023)\\   
   Firm pre-period behavior                        &         &         & Yes     & Yes     &       &         &       &       & Yes\\  
   \midrule
   \emph{Fixed-effects}\\
   Firm                                            &         &         &         &         & Yes   & Yes     &       &       & \\  
   Industry                                        &         &         &         &         &       &         & Yes   & Yes   & Yes\\  
   Experiment start calendar month                 &         &         &         &         &       &         &       & Yes   & Yes\\  
   \midrule
   % \emph{Fit statistics}\\
   Observations                                    & 3,422   & 3,422   & 3,684   & 3,422   & 3,684 & 3,422   & 3,684 & 3,684 & 3,422\\  
   R$^2$                                           & 0.092   & 0.092   & 0.087   & 0.120   & 0.181 & 0.213   & 0.029 & 0.081 & 0.151\\  
   \midrule \midrule
   %                                                 &         &         &         &         &       &         &       &       &  \tabularnewline 
   % \\
\end{tabular}
\par\endgroup
}

    \begin{tablenotes}
       This table reports OLS regressions where the dependent variable is the share of weeks in post-period months 4-6 that an individual treated worker used Copilot. Each treated worker contributes a single observation. Because 80\% of treated individuals used Copilot at least once in this period, most of the variation is on the intensive margin. Firm pre-period behavior controls (Columns 3 and 4) comprise the means of the individual-level regressors aggregated to the firm level. See Appendix \ref{app:outcomes} for variable definitions and Appendix Table \ref{table:horserace_firm_start_month_coefs} for estimated coefficients on firm behaviors and start month indicators. Standard errors are heteroskedasticity-robust.
    \end{tablenotes}
\end{table}

To better understand the importance of the firm in driving adoption, we test how well we can predict workers' use of Copilot by using proxies for the type of work done by individual study participants (pre-experiment work patterns), the type of work done at different firms (firm industry and average work patterns), and how firms distributed their Copilot licenses (the share of a worker's coworkers who had a license). As shown in the first column of Table \ref{table:horserace}, individual pre-experiment activity in Outlook, Teams, and Word explains only 9.2\% of the variation in Copilot adoption for all treated workers in months 4-6. Pre-experiment emails read have a small negative association with the share of weeks an individual used Copilot (the coefficient is per thousand emails), whereas time spent and activity in both Teams and Word are positively associated with use.  Email and Teams meetings are positively correlated, making coefficient interpretation challenging, but the implied usage effect of a standard deviation change in Teams meetings (see Table \ref{table:main_results} for summary statistics) tends to be larger than the implied effect of a standard deviation change in emails read or threads replied to. The share of coworkers that have Copilot, capturing firms' strategy in deploying these early licenses, adds little explanatory power (Column 2).\footnote{The .013 coefficient implies that moving from no coworkers with Copilot to the median of 11.9\% increases the predicted share of Copilot usage weeks by 0.15 percentage points.}  Columns 3 and 4 include firm-level averages of the same pre-period work behavior (coefficients are reported in Appendix Table \ref{table:horserace_firm_start_month_coefs}). They explain 8.7\% of the variation in adoption alone, with explanatory power loading on similar behaviors, and 12\% with the individual-level averages.

Firm fixed effects alone explain nearly twice as much variation in individual workers' Copilot use as workers' pre-period work behavior (comparing Table \ref{table:horserace} Columns 5 and 1). Combining the two sets of predictors explains 21\% of the variation in adoption rates (Column 6). As suggested by Figure \ref{fig:usage_combined}, industry fixed effects (Column 7) explain only 2.9\% of the variation, far less than firm fixed effects. Industry fixed effects and indicators for the month each firm entered the study explains 8\% of the variation (Column 8). However, as shown in Appendix Table \ref{table:horserace_firm_start_month_coefs}, these start month indicators do not follow any predictable pattern.\footnote{If, for example, firms who entered the study later had access to better versions of the Copilot tools that affected uptake, we would expect a monotone pattern, which we do not find. Instead, industry and starting month combinations appear to serve as noisy proxies for firm fixed effects.} The final column of Table \ref{table:horserace} includes individual workers' behavior and all observable firm characteristics except the indicators for firm identity. Together, these factors explain 15\% of the variation in Copilot use, still less than firm fixed effects alone. Overall, these patterns suggest an important role for differences across firms, such as managerial practices or organizational philosophy, that have been highlighted in earlier studies but are unobservable for us.

\section{Direct Effects of Generative AI on Patterns of Work}
\subsection{Methods}
\label{sec:Expdesign2} 

Our main estimation is a difference-in-differences (DiD) approach, which uses pre-experiment data to account for baseline individual heterogeneity and increases precision relative to a simple comparison of means. Using worker $i$ by week $t$ level data, our intent-to-treat (OLS) estimates come from:
\begin{align}
    Y_{it} = \alpha_i + \delta_{m_t} + \gamma_{f,\tau_{ft}} + \beta^{ITT} Z_{i}\cdot\mathbf{1}\{\tau_{ft}\geq0\} + \epsilon_{it} \label{regression}
\end{align}
\noindent
where $f$ is the firm, $\alpha_i$ is an individual fixed effect, $\delta_{m_t}$ captures calendar time (year-by-month) effects, $\gamma_{f,\tau_{ft}}$ are firm-by-event month fixed effects, $Z_{i}$ is the worker's experimental treatment assignment, and $\tau_{ft}$ is the event month for firm $f$. The estimation sample includes only workers in the experiment. Firm-event month fixed effects absorb firm-specific trends and fluctuations in work patterns. Calendar time fixed effects modestly aid precision by further capturing common seasonal trends.\footnote{Both effects are identified because the event-time indicators mark 4-week blocks following each firm's start date, which may fall mid-calendar month.} We cluster standard errors at the individual level.

With this design, the effects of Copilot are identified by differences in the month-to-month changes in outcomes between treated workers and comparison workers at the same firm. The potential biases from a rolling DiD \citep{goodman2021difference} are therefore not a concern. Our baseline estimates omit months 0-3 (weeks 0-12) after each firm's experimental licenses were assigned to focus on the longer-term effects of Copilot use.\footnote{Estimates using all post-treatment weeks are in Appendix Table \ref{table:all_months_iv}.}

We also estimate local average treatment effects by instrumenting a Copilot use indicator with treatment assignment. Instrumenting overcomes issues with non-compliance in both directions: workers who are assigned a Copilot license but do not use the tool and the less common (but non-trivial) case of workers who were assigned to the control group but nonetheless obtain Copilot licenses \citep{duflo2001schooling,hudson2017interpreting}. The outcome equation is \begin{equation}
    Y_{it} = \alpha_i + \delta_{m_t} + \gamma_{f,\tau_{ft}} + \beta^{IV} D_i\cdot\mathbf{1}\{\tau_{ft}\geq0\}  + \epsilon_{it}\label{eqIV}\\
\end{equation}
where $D_i\cdot\mathbf{1}\{\tau_{ft}\geq0\}$ is one in the post-period for individuals that used Copilot in months 4 through 6 and zero otherwise. We estimate the model via two-stage least squares, instrumenting $D_i\cdot\mathbf{1}\{\tau_{ft}\geq0\}$ with $Z_{i}\cdot\mathbf{1}\{\tau_{ft}\geq0\}$.

Because we test multiple hypotheses, we use sharpened $q$-values that account for the increased false discovery rate as the number of tests increases \citep{anderson2008multiple}. The $q$-values (reported as asterisks in tables) can be interpreted like $p$-values after accounting for the total number of hypotheses tested. The number of tests reflect all parameter estimates for Tables \ref{table:main_results} and \ref{table:spillover_table} in the main text, along with the additional outcomes considered in Appendix Table \ref{table:app_results}. 

\subsection{Results}
\label{sec:maineffects}

We focus on two broad sets of outcomes: 1) time use in digital applications, and 2) quantity-related metrics such as the number of meetings attended and documents completed. The time use measures for Outlook and Word are constructed from timestamps associated with user actions and defined as sessions that capture closely-spaced actions. An Outlook session begins when a user opens an email in the reading pane, continues as long as the user records another ``reading" action at least every 15 minutes, and ends when the user exits the reading pane and does not open another email for 15 minutes or more. Word sessions are similarly based on 15 minute gaps in activities (e.g., interactions with a document or series of documents).\footnote{Importantly, we do not have complete measures of workers' actions composing and sending email, so time spent creating emails must be inferred as the time between interactions with the reading pane. The Outlook session metrics were developed in consultation with the Microsoft product team that manages the M365 telemetry data. Patterns of Outlook and Word session time are robust to cutoffs of 10 or 20 minutes rather than 15. Workers may multi-task or switch rapidly between applications such that sessions in different applications overlap. Due to data processing challenges, we do not measure or adjust for this overlap across apps, so session-based measures may overstate the actual time spent on each activity. Details on the construction of all worker behavior measures are in Appendix \ref{app:outcomes}.} The quantity measures are straightforward, as is the time spent attending Teams meetings. Table \ref{table:main_results} contains the main results.

\begin{table}[htbp]
  \caption{Effects of Copilot Provision and Takeup} 
    \label{table:main_results}    \resizebox{\textwidth}{!}{%
\begin{tabular}{>{\arraybackslash}p{6cm} >{\centering\arraybackslash}p{2cm} >{\centering\arraybackslash}p{1.3cm} >{\centering\arraybackslash}p{1.3cm} >{\centering\arraybackslash}p{1.3cm} >{\centering\arraybackslash}p{1.3cm}}
\hline
& (1) & (2) & (3) & (4) & (5) \\
\textbf{Metric} & \textbf{Pre-period mean (SD)} & \textbf{OLS (SE)} & \textbf{IV (SE)} & \textbf{Workers} & $\mathbf{n}$ \\
\hline
Outlook total session time & 11.65 & -1.37** & -2.03** & 6441 & 250145 \\
& (6.81) & (0.13) & (0.19) & & \\
Outlook sessions & 31.67 & -3.56** & -5.30** & 6441 & 250145 \\
& (11.08) & (0.30) & (0.44) & & \\
Email-free work time & 26.18 & 1.48** & 2.20** & 6441 & 250145 \\
& (6.77) & (0.14) & (0.21) & & \\
Outlook out-of-hours session time & 2.19 & -0.33** & -0.49** & 6441 & 250145 \\
& (1.84) & (0.038) & (0.057) & & \\
Emails read & 160.83 & -6.69** & -10.02** & 6387 & 223702 \\
& (117.74) & (1.62) & (2.43) & & \\
Unique email threads replied to & 14.51 & -0.16 & -0.24 & 6441 & 267887 \\
& (14.17) & (0.17) & (0.25) & & \\
Time to reply from email delivery & 16.55 & -0.33 & -0.47 & 6432 & 235121 \\
& (5.34) & (0.17) & (0.24) & & \\
\\[-6pt]
\hdashline\\[-4pt]
Teams meeting time & 5.22 & 0.10 & 0.14 & 6170 & 155003 \\
& (3.65) & (0.054) & (0.076) & & \\
Total Teams meetings attended & 9.93 & 0.24 & 0.34 & 6170 & 253723 \\
& (6.93) & (0.11) & (0.15) & & \\
Share of mtgs. attended early/late (\%) & 30.65 & 0.019 & 0.026 & 6086 & 128660 \\
& (11.94) & (0.27) & (0.37) & & \\
Recurring Teams meeting time & 2.32 & 0.037 & 0.052 & 6170 & 202314 \\
& (1.93) & (0.029) & (0.041) & & \\
\\[-6pt]
\hdashline\\[-4pt]
Word total session time & 1.63 & 0.14 & 0.19 & 2525 & 104913 \\
& (1.58) & (0.065) & (0.089) & & \\
Avg. document time to complete & 186.54 & -3.98 & -5.27 & 2521 & 29683 \\
& (94.20) & (8.96) & (11.86) & & \\
Avg. collaborative document TTC & 287.85 & -48.03 & -60.56 & 1763 & 8700 \\
& (127.78) & (21.50) & (27.17) & & \\
Avg. non-collaborative document TTC & 158.07 & 0.16 & 0.22 & 2506 & 25252 \\
& (89.13) & (9.07) & (12.02) & & \\
Completed documents & 0.76 & 0.029 & 0.039 & 2525 & 77138 \\
& (0.80) & (0.025) & (0.035) & & \\
Completed collaborative documents & 0.14 & 0.011 & 0.015 & 2525 & 77138 \\
& (0.21) & (0.0072) & (0.0098) & & \\
\\[-6pt]
\hdashline\\[-4pt]
Total out-of-hours session time & 2.72 & -0.25** & -0.36** & 6304 & 244944 \\
& (2.16) & (0.050) & (0.075) & & \\
\hline
\multicolumn{6}{l}{\footnotesize Significance: ** $q < 0.01$, * $q < 0.05$} \\
\end{tabular}%
}

    \begin{tablenotes} For each outcome variable, this table shows pre-period mean and cross-worker standard deviations (Column 1), the intent-to-treat coefficients for the effect of Copilot (Column 2), and the LATE coefficients for Copilot adopters (Column 3). All time metrics are in hours. There are some differences in worker counts across metrics due to differences in data collection and storage policies across sources, averages not being meaningful when counts are zero, and our exclusion of workers with low/zero pre-period activity for Word and Teams. See Appendix \ref{sec:dataapp} for details. Share of meetings attended late/early refers to the share of Teams meetings that a worker joined more than 5 minutes after the meeting started or left more than 5 minutes before it ended. We assign documents to a worker if they were the ``primary editor," making the  most edits to the document. Time to complete (TTC) and completed document counts are calculated only for documents where the target worker is the primary editor. Standard errors are clustered at the individual level.  Asterisks reference sharpened $q$-values that adjust for the false discovery rate, as in \citet{anderson2008multiple}.
    %A worker is a primary editor if they make the most edits on a document. 
    \end{tablenotes}
\end{table}

\subsubsection{Effects on Time Use}

The most prominent impact of Copilot on worker behavior that we observe involves changes in how workers manage email. In the pre-period, the average worker spent 11.7 hours per week reading and responding to emails. Our intent-to-treat (LATE) estimates show workers with access to Copilot spent an average of 1.4 (2) fewer hours per week in email sessions over post-period months 4-6, a 12\% (17\%) decline from their pre-period average. Workers with access to Copilot also consolidated their email work, reducing their count of Outlook sessions per week by 11\% (17\%), while opening up almost 1.5 hours per week of time in ``non-email" blocks of 30 minutes or more during the regular workday with no Outlook activity (LATE = 2.2 hours). These email-free blocks capture longer periods where workers may be able to focus on other projects without interruptions from email \citep{jackson2001cost}. 

We find no similar shifts in the time workers spent in Teams meetings or working in Word, despite the fact that workers regularly used Copilot through these applications. The ITT results rule out effect sizes outside of the interval between -0.01 and 0.21 hours for Teams meetings relative to a mean of 5.22 hours per week. We estimate an increase in Word session time of 0.14 (LATE = 0.19) hours per week, an 8.6\% increase from a small pre-period mean, but this effect just misses statistical significance using the sharpened $q$-values. The smaller sample of regular Word users leaves us somewhat under-powered to detect changes in document-related work.\footnote{We believe differences in Word use reflect the range of jobs done by our sample of workers. See Word usage by occupation in Appendix \ref{app:jobtitles}.} 

\begin{figure}[htbp]
    \centering
    \begin{subfigure}{\textwidth}
        \centering
        \caption{Time in Outlook
        \label{fig:event_outlook}}
        \includegraphics[width=\textwidth]{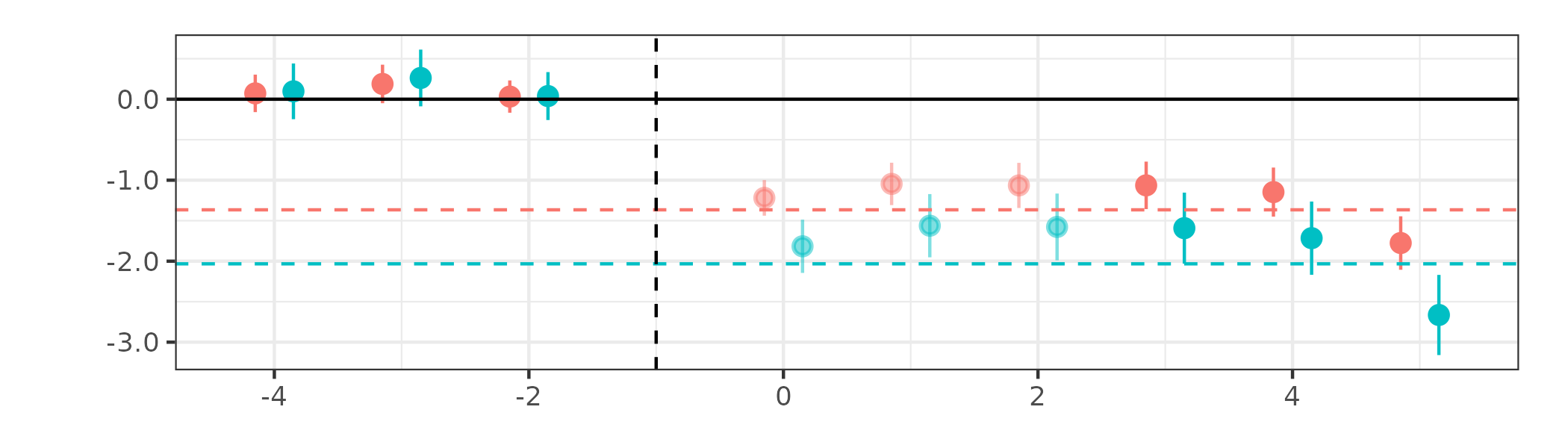}
    \end{subfigure}
    \\
    \begin{subfigure}{\textwidth}
        \centering
       \caption{Time in Teams Meetings
        \label{fig:event_meetings}}
        \includegraphics[width=\textwidth]{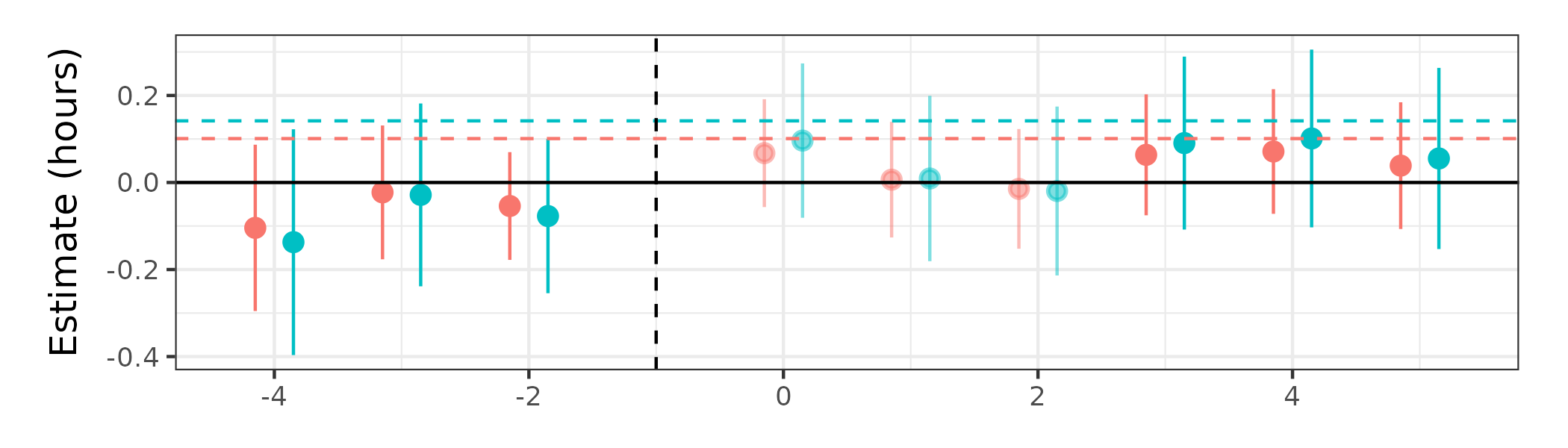}
    \end{subfigure}
    \\
    \begin{subfigure}{\textwidth}
        \centering
        \caption{Time in Word %- Across Firms
        \label{fig:event_word}}
        \includegraphics[width=\textwidth]{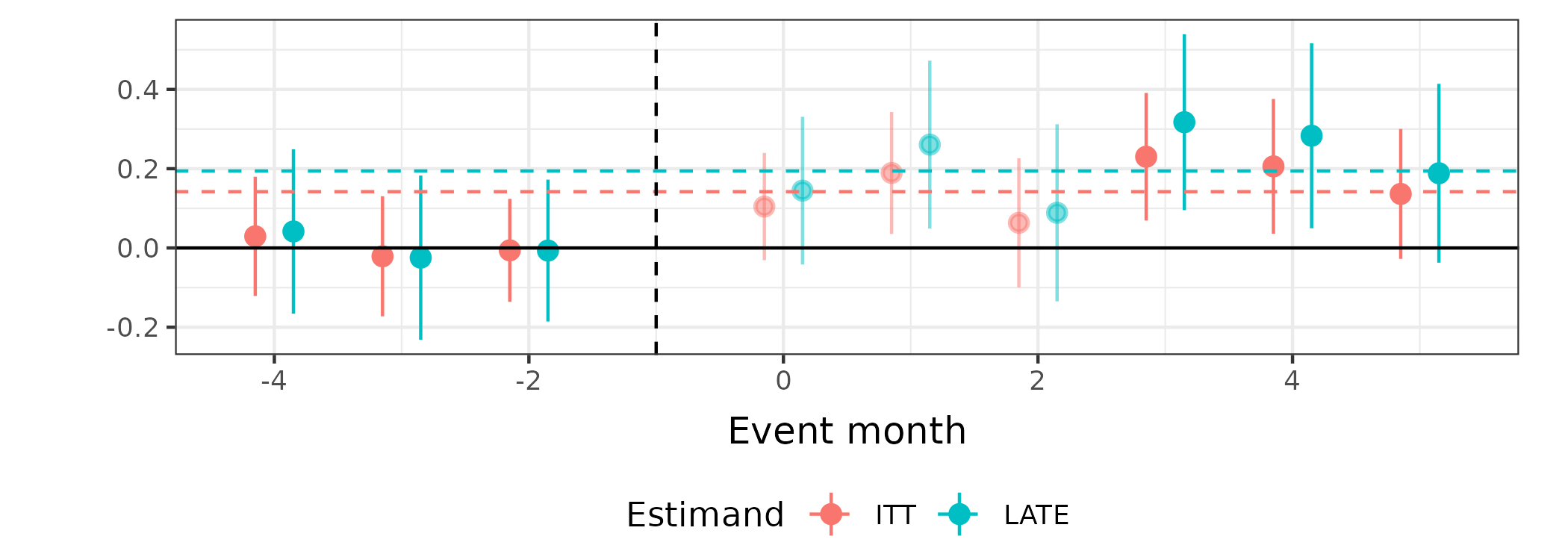}
    \end{subfigure}

    \hfill
    \caption{Event Study Plots for the Effect of Copilot on Time Allocation\label{fig:event_studies}}
    \begin{tablenotes}
        These plots show coefficients from event-study versions of the regressions defined in Equations \eqref{regression} and \eqref{eqIV}. For the ITT estimates, we estimate event month-specific treatment effects, $\beta_\tau$ using  $Y_{it} = \alpha_i + \delta_{m_t} + \gamma_{f,\tau_{ft}} + \sum_{k \neq -1} \beta_\tau Z_{i} \cdot \mathbf{1}\{\tau_{ft} = k\} + \epsilon_{it}.$ To extend the IV to a multi-period event study, we interact $D_{i}$ with event month and instrument with interactions of $Z_{i}$ and event month. While outcomes are measured on a weekly basis, time-varying effects are at a monthly frequency to reduce noise. Vertical lines show 95\% confidence intervals based on individual-clustered errors. Horizontal dashed lines give the average estimated effect from Table \ref{table:main_results}. These plots show treatment effects by event time, which pools calendar months across firms in our staggered adoption setting. 
    \end{tablenotes}
\end{figure}

Across applications, the last row of the table shows workers spent a quarter of an hour less per week working outside of their modal work hours, a 9\% reduction, with somewhat larger time savings in after-hours email work. Together, these results suggest that workers used some of their time saved on email to shorten their work day, but much of the time saved appears to have shifted to activities we cannot measure. The dynamic difference-in-differences results in Figure \ref{fig:event_studies} illustrate that the effects on Outlook time are relatively stable over the full post-period, with persistent effects that are apparent even in the first month after workers had access to the tool. The Teams and Word results remain close to zero in all months. 

Although our treatment comprises access to, or use of, Copilot through any Office application, the Outlook time savings likely come from Copilot use in Outlook rather than cross-application effects.  While transcription tools in Teams Copilot could, in principle, reduce the need to write follow-up emails, reduced form regressions of outcomes on application-specific Copilot use (see Appendix Table \ref{tab:indiv_copilot_redform}) do not find much correlation between Teams Copilot use and changes in outcomes for email or document completion after controlling for use of other Copilots.\footnote{Note that Copilot use is strongly correlated across applications (see Appendix Table \ref{tab:use_correl}), limiting our power to detect cross-application relationships.}

Given the differences in Copilot adoption in Table \ref{table:horserace}, a natural question is whether treatment effects also vary by group.  We use the machine learning procedures in \citet{chernozhukov2025heteffects} to evaluate heterogeneity in conditional average treatment effects (CATEs) across a high-dimensional set of covariates.  We cannot reject the null of 0 treatment effect heterogeneity for total Outlook session time when we allow the model to choose which dimensions of potential heterogeneity matter via elastic net or a neural networks approach (see Appendix Table \ref{table:heterogeneity} for details and estimates).  Group average treatment effects for the top and bottom quintiles of workers by CATE are only modestly different than the full average treatment effects in Table \ref{table:main_results} (-1.65 and -1.18 hours for the most and least affected workers) and we cannot reject that treatment effects for these two groups are equal.

\subsubsection{Effects on Task Quantity and Type}

In addition to consolidating their email work into fewer sessions per week, we find workers with Copilot also generated time savings by reading 4\% (LATE = 6\%) fewer emails each week (Table \ref{table:main_results})\footnote{Our measure counts unique emails read each week. If a worker revisited an email, those actions would count as email reads in each week. If a worker only used Copilot to summarize information, that would not count as an email read.} with no increase in their time spent per email (in Appendix Table \ref{table:app_results} we rule out increases larger than 1.2\%). Table \ref{table:main_results} shows no significant changes in the number of unique threads replied to or in the time from delivery to reply (we can rule out effect sizes greater than 3.4\% of the mean in the ITT estimates for email threads and greater than 4.1\% of the mean for reductions in time to reply to emails), suggesting that workers were not ignoring their inboxes or missing emails that required a reply. The number of emails per thread (Appendix Table \ref{table:app_results}) did not change, suggesting that Copilot may have been used for email summaries but with limited effects on the mix of emails or on thread efficiency. Given the estimated time savings, workers' responses appear to reflect their ability to take less time to do roughly the same quantity of email work.  

The null effects of Copilot on time spent in Teams meetings or Word document writing could reflect offsetting responses to a productivity-enhancing tool: workers complete the same work in less time but use the time saved to do more of that kind of work. These competing shifts are akin to the income and substitution effects generated by a change in price --- in this case, the amount of time required to complete a task. If there were offsetting or counterbalancing effects, we would expect to see quantity responses, such as an increase in the number of meetings or documents. We might also expect  shifts in the composition of tasks, like observing more short meetings or fewer recurring meetings. However, we find no evidence for these kinds of changes for Teams tasks in Table \ref{table:main_results} (we can rule out increases in the number of meetings greater than 4.6\%).\footnote{As shown in Appendix Table \ref{table:app_results}, we find one statistically significant compositional change: a small (0.06 to 0.085 per week) increase in the count of short meetings with many participants. \citet{rogelberg2019surprising} finds that large meetings are most effective for information transmission rather than ideation or problem solving. Copilot may have swayed busy workers who previously skipped some of these meetings to attend instead because they could transcribe and catch up later, even if multitasking in the moment.} 

The patterns in the third panel of Table \ref{table:main_results} are consistent with these two counterbalancing effects, but the estimates are imprecise and the point estimates are small. Regular Word users with Copilot completed their documents more quickly, consistent with \citet{noy2023experimental}, but the ITT point estimate is a savings of only 2\% of the pre-period mean, and we cannot rule out that document time to complete increased by as much as 7\%.\footnote{Note that the time to complete a document is distinct from the total time spent in Word. Time to complete is based on the elapsed time from when a document is created to its approximate end (when 90\% of all edits are done), whereas Word session time measures the actual time users spend in Word.} We estimate that Copilot use also slightly increased the total Word documents authored by the studied workers (we can rule out increases larger than 11\% of the pre-period mean and cannot rule out no change). Time to complete fell the most for collaborative documents (over 2 days, or a 17\% reduction), however even these effects are not statistically significant using the sharpened $q$-values.\footnote{A document is defined as collaborative if a worker was the primary editor but at least one other worker contributed. Non-collaborative documents are individual.} Any time savings are not accompanied by an increase in collaborative document production or the share of authored documents that include collaborators.

\section{Work Quality, Spillovers, Group Effects, and Co-Invention}

\begin{table}[htbp]
  \caption{Spillover Effects of Copilot} 
    \label{table:spillover_table}    \resizebox{\textwidth}{!}{%
\begin{tabular}{>{\arraybackslash}p{6cm} >{\centering\arraybackslash}p{3cm} >{\centering\arraybackslash}p{3cm} >{\centering\arraybackslash}p{3cm}}
\hline
& (1) & (2) & (3) \\
\textbf{Metric} & \textbf{Direct treatment (SE)} & \textbf{Teammate treatment (SE)} & \textbf{Direct $\times$ Teammate treatment (SE)} \\
\hline
Outlook total session time & -1.26** & -0.37 & -0.29 \\
& (0.16) & (0.17) & (0.26) \\
Outlook sessions & -3.32** & -0.54 & -0.68 \\
& (0.36) & (0.38) & (0.60) \\
Email-free work time & 1.38** & 0.41 & 0.28 \\
& (0.17) & (0.18) & (0.28) \\
Outlook out-of-hours session time & -0.28** & -0.060 & -0.13 \\
& (0.047) & (0.054) & (0.077) \\
Emails read & -7.09** & -3.59 & 1.00 \\
& (1.95) & (2.34) & (3.36) \\
Unique email threads replied to & -0.35 & -0.73* & 0.52 \\
& (0.20) & (0.26) & (0.36) \\
Time to reply from email delivery & -0.16 & 0.097 & -0.47 \\
& (0.21) & (0.26) & (0.34) \\
\\[-6pt]
\hdashline\\[-4pt]
Teams meeting time & 0.17 & -0.015 & -0.19 \\
& (0.069) & (0.084) & (0.11) \\
Total Teams meetings attended & 0.35 & -0.35 & -0.29 \\
& (0.14) & (0.17) & (0.23) \\
Share of mtgs. attended early/late (\%) & 0.039 & 0.38 & -0.059 \\
& (0.36) & (0.41) & (0.55) \\
Recurring Teams meeting time & 0.077 & -0.050 & -0.11 \\
& (0.036) & (0.048) & (0.061) \\
\\[-6pt]
\hdashline\\[-4pt]
Word total session time & 0.084 & -0.066 & 0.16 \\
& (0.083) & (0.10) & (0.13) \\
Avg. document time to complete & -11.86 & -11.06 & 21.07 \\
& (11.03) & (13.65) & (18.26) \\
Avg. collaborative document TTC & -67.20 & -26.75 & 43.06 \\
& (29.03) & (30.34) & (43.94) \\
Avg. non-collaborative document TTC & -0.63 & 4.81 & 2.72 \\
& (11.05) & (13.71) & (18.53) \\
Completed documents & 0.038 & 0.053 & -0.025 \\
& (0.032) & (0.040) & (0.053) \\
Completed collaborative documents & 0.011 & 0.0028 & -0.00079 \\
& (0.0087) & (0.011) & (0.015) \\
\\[-6pt]
\hdashline\\[-4pt]
Total out-of-hours session time & -0.22** & 0.034 & -0.074 \\
& (0.064) & (0.074) & (0.10) \\
\hline
\end{tabular}%
}

    \begin{tablenotes}This table shows estimates resulting from a triply-interacted version of the main OLS regression specification, where the three treatments of interest are being in the treatment group, having a close coworker who received Copilot through the experiment, and being in the treatment group while also having a treated coworker. Reported coefficients are the interactions of these terms with the post-period indicator. Asterisks reference sharpened $q$-values that adjust for the false discovery rate, as in \citet{anderson2008multiple}.
    \end{tablenotes}
\end{table}

The patterns in Table \ref{table:main_results} suggest that workers with Copilot took less time to do roughly the same quantity of email work. Without observing the content of the emails they write, we cannot say definitively whether this time savings came at the cost of reduced work quality. However, we can provide suggestive evidence by considering spillover effects on the colleagues of treated workers.
As discussed in Appendix \ref{app:coworkers}, we identify close colleagues as the coworkers one interacts with most frequently through emails, meetings, and joint writing. If treated workers were inefficiently reducing time spent on email or sending lower quality replies generated by Copilot, we might expect to see their colleagues spending more time in email to pick up the slack. Alternatively, treated workers may have established effective new practices that also benefited their teams. 

To test these channels, we estimate a reduced-form intent-to-treat model that captures potential spillovers.  In doing so, we are also able to account for potential violations of the Stable Unit Treatment Value Assumption (SUTVA) because of spillovers to the control group and to consider whether Copilot provision to a close coworker changes treatment effects for the focal worker.  The regression is

\begin{equation}
\begin{split}
    Y_{it} = \alpha_i + \delta_{m_t} + \gamma_{f,\tau_{ft}} + \beta^{ITT} Z_{i}\cdot\mathbf{1}\{\tau_{ft}\geq0\}+\beta^{Close}Z_{Close}\cdot\mathbf{1}\{\tau_{ft}\geq0\} \\
    +\beta^{ITT\cdot Close}Z_i\cdot Z_{Close}\cdot\mathbf{1}\{\tau_{ft}\geq0\}+ \epsilon_{it} 
    \end{split}
    \label{saturated}
\end{equation}
\noindent 
where the model mimics equation (\ref{regression}), adding the indicator for having a close connection who is offered Copilot through the experiment, $Z_{Close}$, and an interaction of the focal worker's treatment and $Z_{Close}$. As shown in Appendix Figure \ref{fig:coworker_share}, 30\% of studied workers have at least one treated coworker during the post-period.\footnote{Workers may also have close colleagues who received one of the firms' discretionary (i.e. not randomly assigned through the experiment) Copilot licenses. Under our data use agreements, we are unable to track this Copilot access for workers outside the study.}

The first column of Table \ref{table:spillover_table} shows the main treatment effect estimates when we control for potential spillovers to the control group through $Z_{Close}$, a strategy that \cite{de2012effects} and \cite{espinosa2022training} use to assess the impact of potential SUTVA violations.  The point estimates are very similar to the reduced form estimates in Table \ref{regression}, have the same patterns of statistical significance based on the sharpened q-values, and generally fall within the confidence intervals of the original estimates.

It should not be surprising, given the stability of estimates for the focal worker, that the results in column two for having a close co-worker treated are generally much smaller in magnitude than the main effects. The spillover effects on coworkers' email activity generally have the same sign as the main effects. However, adjusting for multiple hypothesis testing, the only statistically significant estimate is on unique email threads replied to, which falls.  Overall, these spillover estimates suggest that treated workers were using Copilot to reduce email time in efficient ways, with modest benefits for their coworkers.  This provides suggestive evidence that their gains did not come at the expense of quality.

In keeping with the largely null direct effects of Copilot for meetings or writing tasks, there are no spillover effects of Copilot access on connected coworkers' Teams or Word activity. These consistent differences across applications may reflect how easily individual workers can adjust their work to take advantage of the new tool. Email is solitary and each worker is free to develop her own process for inbox management. In contrast, shifting what work gets done in meetings, how long they last, or who takes primary responsibility for writing documents requires coordinating with colleagues and agreeing on new norms. This pattern is fitting for the early adoption stage we study, where individual workers were exploring the capacities of a new tool, but the complementary process innovations and organizational restructuring that drive larger changes in workflow had not yet taken place \citep{bresnahan1996technical, brynjolfsson2000beyond}.

In the spirit of the literature on complementary innovation and process restructuring, we test whether there are different effects when multiple coworkers get access to the tool. The third column of Table \ref{table:spillover_table} shows the triply interacted effects of $Z_i$, $Z_{Close}$, and $\mathbf{1}\{\tau_{ft}\geq0\}$. The total effect of being treated oneself while having at least one treated close colleague is given by the sum across the three columns. The point estimates suggest some modest benefits to gaining access to the new tool along with colleagues, treated workers with treated colleagues save 50\% more time on email, but again these effects are imprecisely estimated and, with the sharpened q values, we cannot rule out zero differences in effect.\footnote{The specification here is a 0-1 indicator for whether any close coworker is treated, while a saturation design would account for the share of workers on a team that are treated.  Using that approach, reported in Appendix Table \ref{table:split_results_teammates}, we also detect little difference in treatment effects by whether our studied workers had an above or below median share (11.9\%) of close coworkers with access to Copilot. However, even our ``high saturation" teams had, on average, only 39\% of coworkers with access to Copilot, so we may simply be unable to detect team-level effects from this experimental design.} These results suggest that the co-creation of work and process restructuring that the literature identifies as a pre-condition for effective technology adoption did not seem to take place under the individual-level rollout of the tool that our experiment assesses.

\section{Conclusion}
\label{sec:conclusion}

This paper offers a portrait of the first steps of adoption of a new technology. Historically, the main productivity gains from innovative technologies came only with the complementary process innovations that shift the nature of work around these new capacities.  These institutional changes take time. Among recent technological innovations, generative AI is unique because it does not require new equipment or hardware to use, allowing individual workers to bring generative AI to work even without their employers' support. As \citet{blandin2024rapid} point out, the discrepancy between the high rates of generative AI use reported in surveys of workers, such as theirs, and the lower rates reported in surveys of firms, as in \citet{bonney2024tracking}, suggests that workers use these tools even before their employers develop formal adoption strategies.

This study provides indications of how this individual use of generative AI shapes the patterns of work for knowledge workers in large firms. Large firms are often among the earliest adopters of digital technologies, presumably because of their ability to capture benefits \citep{mcelheran2024ai}. We study workers during the first year in which generative AI tools gained broad use. While their colleagues may have been using other generative AI tools, the workers we studied were often the only members of their team with access to Copilot during the early roll-out period. Although we cannot measure productivity, the changing behaviors we see are consistent with workers independently exploring these new tools and saving time on individual tasks. Further research is needed to assess how co-inventions or team- or firm-level transformations may lead to broader changes, such as shifts in what tasks workers do.

%\pagebreak
\FloatBarrier
%\bibliographystyle{aea}
%\bibliography{Bibliography}

% Bibstyle aea.bst version 2009.05.20

\clearpage
\newpage
\setcounter{table}{0}
\setcounter{figure}{0}
\setcounter{section}{0}

\renewcommand\thesection{\Alph{section}}
\renewcommand\thetable{\thesection.\arabic{table}}
\renewcommand\thefigure{\thesection.\arabic{figure}}
\renewcommand{\topfraction}{.9}
\renewcommand{\floatpagefraction}{.9}

\part*{Supplemental Appendix}

\section{Experiment Details \label{app:exp}}

The experiment was overseen by Microsoft's marketing group in collaboration with firms. Participating firms chose how they wanted to select workers for the study and were responsible for all communication with participating workers, including recruiting and obtaining consent. Firms shared a list of anonymized identifiers for the participating workers with the experiment organizers at Microsoft. The Microsoft side randomized treatment assignment within this list of workers, shared the assignment back to the firm, and used the anonymized identifiers to track telemetry data for the studied workers. The authors of this paper received permission from Microsoft marketing and the participating firms to use the data collected through the experiment for research.

Participating firms were told that they would need to allocate at least 50 of their 300 Copilot licenses at random and would need to recruit at least twice as many workers to join the experiment in order to populate treatment and control groups. For example, a firm that chose to allocate 150 licenses to the experiment would need to recruit 300 workers. One firm had more than 500 workers participate in the study (that is, nearly all their initial package of licenses were allocated through the study), while another defied the suggested minimum and had only 68 workers participate. Across firms, the average number of treated workers was 55.8, with a standard deviation of 33.1.  

The incentive for firms to participate in the experiment was that Microsoft agreed to provide participating firms with early access to the research findings and a firm-level analysis of Copilot usage rates among the study participants and some preliminary effect estimates of Copilot on outcomes such as emails read. Firms received the initial report 2 to 4 months after starting the experiment, usually before the end of the studied period. Some firms may have tried to encourage additional adoption due to low adoption numbers in the report, but we do not see systematic changes around these read-outs in the event studies. Many firms also had access to data on Copilot usage for their own workers through other Microsoft products outside of this experiment.

\begin{figure}[ht]
\label{fig:firms}
    \centering
    
    %\begin{minipage}
         \begin{subfigure}{0.48\textwidth}
        \centering
        
        \includegraphics[width=\textwidth]{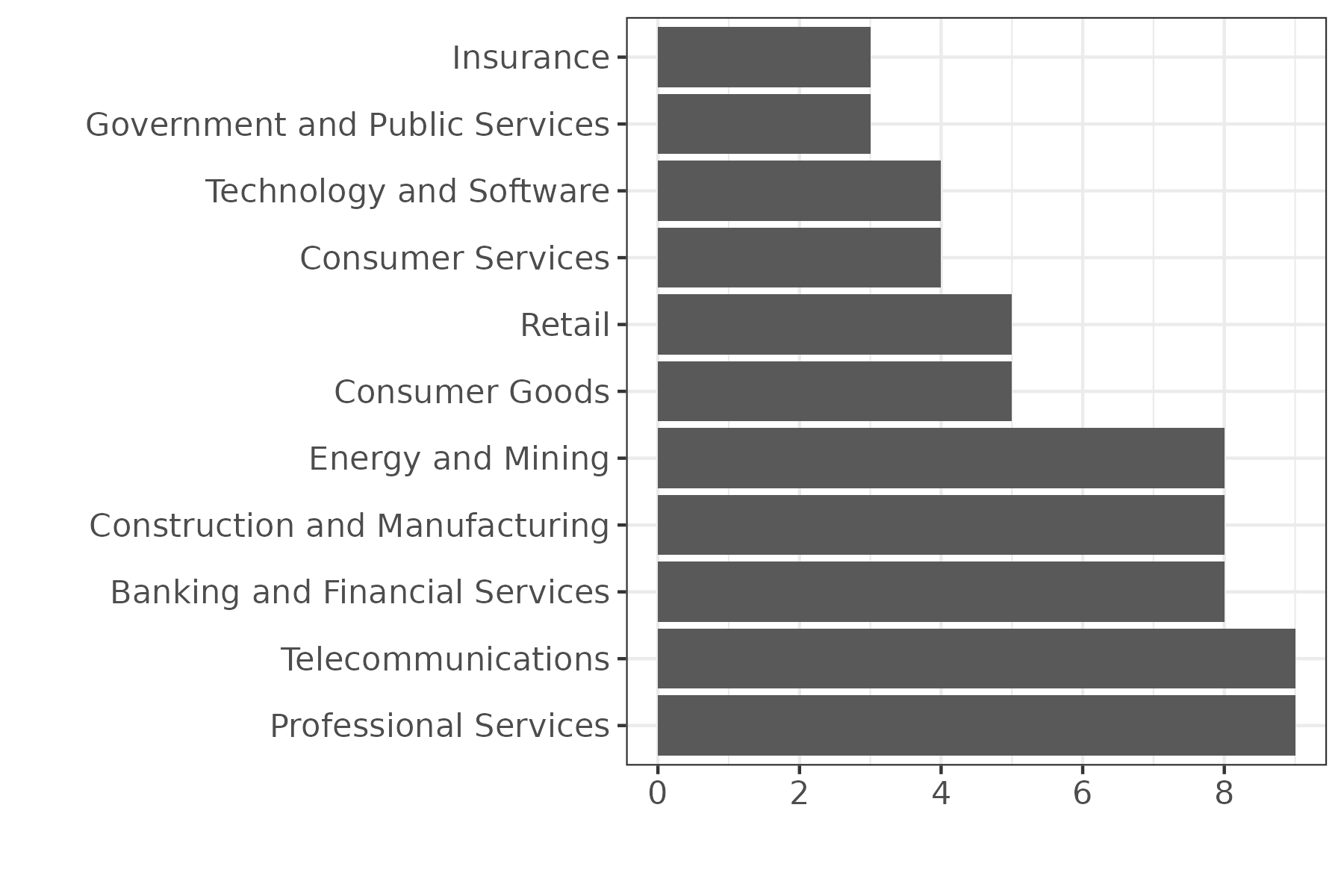}
        \caption{Industry Mix  \label{fig:firm_ind}}
     \end{subfigure}%\end{minipage}
          \hfill
    \begin{subfigure}{0.48\textwidth}
        \centering
        
        \includegraphics[width=\textwidth]{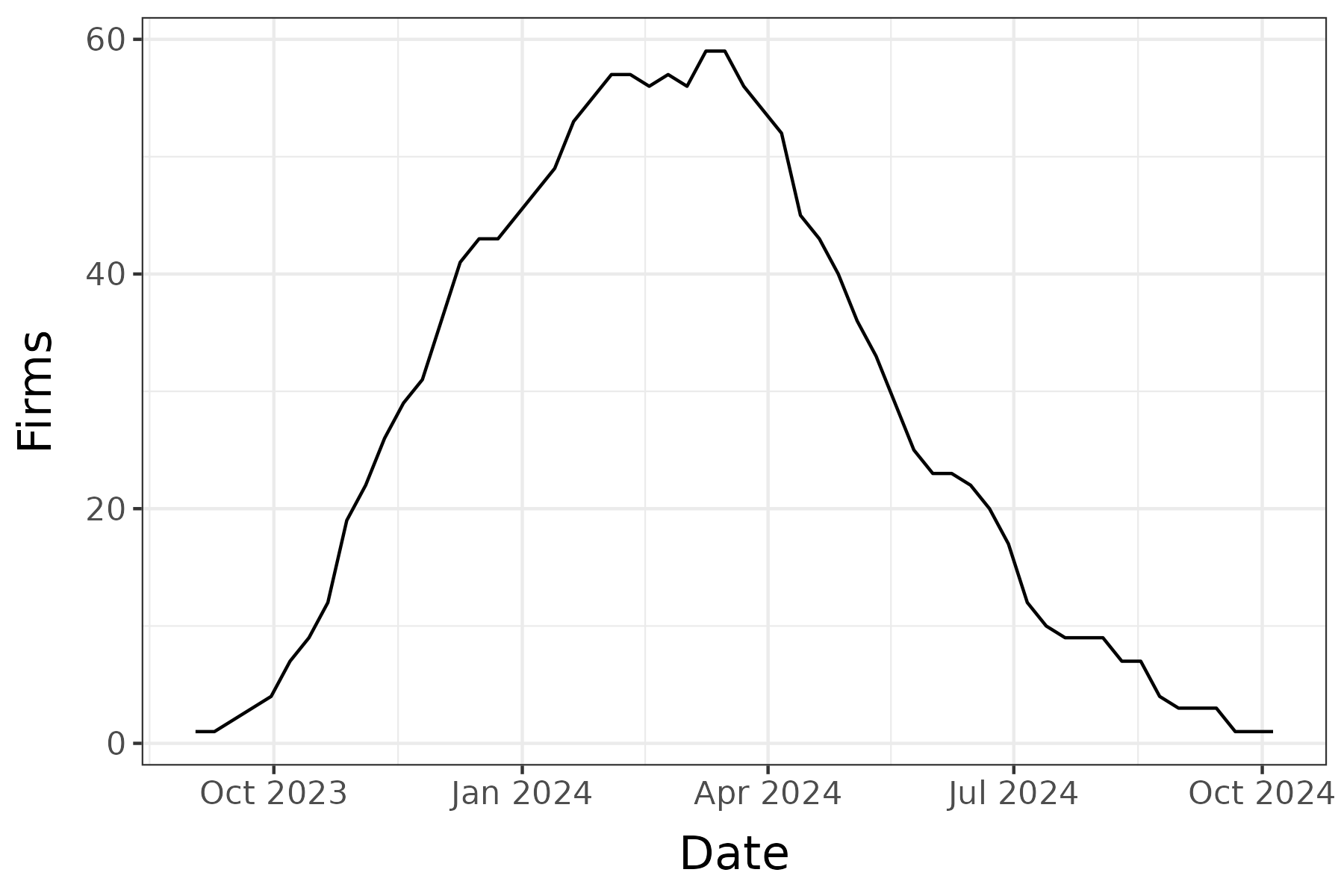}
        \caption{ Over Time\label{fig:firm_time}}
     \end{subfigure}%\end{minipage}
     \caption{Participating Firms}
     \begin{tablenotes}
         Figure (a) shows the number of participating firms in each industry. Figure (b) shows the number of firms active in the experiment over time. 
     \end{tablenotes}
\end{figure}

Our analysis sample includes 66 firms spanning a range of industries tabulated in Figure \ref{fig:firm_ind}. Two additional firms agreed to participate in the study but did not follow through with randomization; these firms are excluded from the analysis sample. The experiment includes workers from many countries, with the majority based in the United States or Europe. 

For all firms, our data sample begins on May 1, 2023, providing at least 4 months of data before any Copilot licenses were allocated. We measure the start of treatment for each firm using telemetry data that captures when M365 Copilot licenses were activated for participating workers. When firms alerted Microsoft of the date they allocated licenses, this start date generally aligns with our measured date, but not all firms provided this information. Most firms allocated all experimental Copilot licenses at once, but a few allocated licenses in batches. In a few cases, we observe one or two participating workers receiving Copilot licenses well before the other participating workers at their firm or the stated start date (likely imperfect compliance with the experiment or possibly a data error). For consistency, we set the start date as the day when 25\% of the licenses for the experiment were allocated. The earliest start date was in September 2023, and the latest in April 2024. Figure \ref{fig:firm_time} illustrates the number of firms for whom we are collecting post-activation data over time.  We collected data on the workers at each firm for a maximum of 24 weeks after licenses were allocated. 24 weeks is also the modal length of the post period across firms. Five firms have shorter treatment periods because they chose to withdraw from the study before 24 weeks. In all analyses, we drop outcomes for the weeks ending December 29, 2023 and January 5, 2024 as work activity was low across all workers during this holiday period.

\FloatBarrier
\section{Data Details \label{sec:dataapp}}
\setcounter{table}{0}
\setcounter{figure}{0}

We analyzed hundreds of millions of Microsoft 365 usage signals (telemetry), for both the treated and designated control workers, from  product telemetry data that Microsoft collects. To maintain workers' and firms' privacy, we do not observe anything about the content produced by these workers, so our focus is on how workers allocate their time and interact with each other rather than on the quality of their work. 

One limitation of this study is that we only observe activity within the Microsoft Office products we can track. We are, for example, measuring time spent in meetings through Teams, not time spent in meetings overall. Fortunately, the firms invited to participate in the study are all heavy users of Microsoft products. The vast majority of workers in our sample used Outlook regularly (6,485 out of 7,137) in the pre-period months, and we suspect we are capturing a large share of work-related email activity. Our baseline sample for email outcomes includes 6,441 workers for whom we observed email sessions in the experiment period and any pre-period Outlook use. Our sample differs for some Outlook metrics for reasons we detail in the next subsection.

In contrast, many sample workers show almost no Word activity during our pre-period months. Based on conversations with participating firms and the patterns of application use by occupation in Table \ref{table:occupation_table}, we think this mostly reflects roles that do not involve document creation, though some low activity could reflect use of a different word processing tool. We do not expect (and do not observe) that individual access to Copilot would shift many of these workers towards using Word. If it did, that change from zero to positive activity would have a very different interpretation than the intensive margin behavioral changes of workers with recorded Word activity in the pre-period. We therefore restrict all Word-related analyses to workers with moderate use in the pre-period, indicated by above-median time spent in Word (0.25 hours/week) and above-median number of documents originated as the `primary editor' (0.16 documents/week). Our main Word sample includes 2,525 workers who meet these criteria.

In the case of meetings, workers with near-zero pre-period use of Teams are concentrated in 10 firms, all of whom report primarily using non-Microsoft applications for video meetings. As with document creation, we are unable to establish reliable pre-period baseline meeting activity for these workers. In this case, we believe this largely reflects our inability to measure work meetings rather than true zeros. Any Copilot-induced changes in Teams activity is likely to reflect substitution to Teams rather than true changes in work meetings. We therefore drop workers at these 10 firms for all meeting outcome analyses. Our main Teams sample includes 6,170 workers in the remaining firms.

\subsection{Outcome Metrics\label{app:outcomes}}
We construct weekly metrics capturing each worker's email, document editing, and meeting behaviors. Metrics are winsorized to the 99th percentile to address telemetry errors and privacy concerns. In the following subsections we define each behavior measured used in the paper. We also note outcome-specific changes in the worker sample or worker-week observation count. Table \ref{table:balance_table} shows the balance tests for pre-experiment differences between the treatment and control groups for all outcomes we use in the analysis. 

\subsubsection{Outlook}
For email, we aggregate the signals of various `reading actions' users take into `Outlook sessions:' blocks of time where a user doesn't go more than 15 minutes without reading an email.\footnote{Read actions are the most commonly and frequently triggered action when people are interacting with Outlook and hence provide the best summary estimate of an Outlook session.} Session time begins at the first eligible moment outside a prior session when a user opens an email in the reading pane. It ends either when the user exits the reading pane and doesn't open an email for at least 15 minutes, with the exact end of the session being the earlier of the user's exit or 15 minutes after the reading action. The user's ``working day'' is defined as the 8 hours in which they were most active in Outlook, during two fixed months of the pre-period immediately prior to the experiment (June and July 2023).

Table \ref{table:main_results} reports the number of unique workers and worker-week observations included in the treatment effect regressions for each outcome (defined below). `Unique email threads replied to' shows the maximum observations we can include. Emails read and all session time metrics have slightly fewer work-week observations (and for emails read fewer workers included) because of an intermittent telemetry outage during the first half of 2024 that prevented us from collecting read action data for some workers in some weeks. `Time to reply from email delivery' has fewer observations because it can only be defined in weeks where the worker replied to at least one email.
\begin{itemize}[noitemsep]
    \bitem{Outlook total session time} The time spent aggregated across all Outlook sessions in a week, measured in hours.
    \bitem{Outlook sessions} The number of Outlook sessions in a week.
    \bitem{Email-free work time} The total time spent during an individual's working day in non-Outlook blocks at least 30 minutes long, measured in hours.
    \bitem{Outlook out-of-hours session time} The subset of Outlook session time that occurred outside the individual's working day - measured in hours.
    \bitem{Unique email threads replied to} The number of unique email threads/conversations among emails that are received that week and replied to within 7 days.
    \bitem{Time to reply from email delivery} The average length of time it takes to reply to an email among emails received that week and replied to within 7 days, measured in hours. 
    \bitem{Emails read} The number of distinct emails read within a week.
    \bitem{Total emails in threads replied to} The total number of emails in the threads/conversations among emails that are received that week and replied to within 7 days.
    \bitem{Average email read duration} The average time spent interacting with (reading) an email, measured in seconds.
    \bitem{Number of email recipients} The average number of recipients on a worker's sent email, observed over a fixed pre-period (April-July 2023).
\end{itemize}

 \subsubsection{Teams}
 All time metrics are measured in hours. Referring again to the last two columns of Table \ref{table:main_results}, `Total Teams meetings attended' shows the maximum workers and observations for our meeting metrics. Teams data have particularly strict deletion windows (that is, Microsoft only stores individual telemetry data for a short period). Teams meeting time and recurring meeting time include the same number of workers, but fewer observations because we were only able to capture pre-period data starting later in the summer of 2023 (the start date varies by firm based on when they joined the study). We can only define the share of meetings attended early or late in weeks when workers attended at least one meeting.
 
 \begin{itemize}[noitemsep]
     \bitem{Teams meeting time} The time spent in scheduled Teams meetings (excluding spontaneous calls)  in a week.
      \bitem{Total Teams meetings attended} The number of Teams meetings attended in a week.
     \bitem{Teams out-of-hours meeting time} The subset of Teams meeting time that occurred outside the individual's working day (see above).
     \bitem{Share of meetings attended early/late (\%)} The share of scheduled Teams meetings joined either $\geq$5 minutes after the Teams meeting was started or left $\geq$5 minutes before the meeting was ended.
    \bitem{Total recurring meetings} The number of Teams meetings that were marked recurring attended in a week. We define recurring meetings as meetings that are scheduled at the same time for the same duration week over week, regardless of whether all the same people attend in each week.
     \bitem{Recurring Teams meetings time} The time spent in recurring meetings.
     \bitem{Big, short meetings} Teams meetings of length less than 1 hour with over 8 people.
     \bitem{Big, long meetings} Teams meetings of length 1 hour or more with over 8 people.
     \bitem{Small, short meetings} Teams meetings of length less than 1 hour with 8 or fewer people.
     \bitem{Small, long meetings} Teams meetings of length 1 hour or more with 8 or fewer people.
     \bitem{Number of unique people met 1-1} The total number of people met in a 1-1 meeting, observed over a fixed pre-period (April-July 2023).
 \end{itemize}
  Note that the cutoff of eight people for small meetings and one hour for short meetings come from the Viva Insights product, which reports these categories. Viva based these categories on research such as \cite{rogelberg2019surprising}, which refers to the 8-18-1800 rule suggesting that ``if you are trying to solve a problem or make a decision, keep the meeting to eight people or less.'' 
\subsubsection{Word} 
    Time use in Word is also measured by session time but is calculated based on intentional app actions (typing or changes to a document) rather than emails read individually or in sequence.  Across metrics, a worker is a primary editor if they make the most edits on a document; a document is complete when 90\% of the total edits to the document have been completed. Some documents show stray edits long after most active work is done, which may be accidental. This 90\% cutoff aligns most often with other indicators of completion such as attaching to an email or printing to pdf. 

    All outcomes in Table \ref{table:main_results} except `Word total session time' come from an underlying telemetry source that was subject to a stricter data deletion window. We collect micro data starting in June 2023, but our first week of these constructed outcomes is the first week of July 2023. Because we measure time to complete in the week the document was completed, we leave a few ``runway" weeks so we can capture the first edits to documents that were completed in July. The time to complete outcomes all have fewer observations because they are only defined in worker-weeks where the worker completed at least one document (or one solo or collaborative document). All time metrics are measured in hours.
   \begin{itemize}[noitemsep]
     \bitem{Word total session time} The time spent in Word sessions, identified as blocks of actions with $<15$ minute gaps.
     \bitem{Avg. document time to complete} The average time between the creation (first edit) and completion of a document, for documents completed that week and where the target worker is the primary editor. 
     \bitem{Avg. collaborative document time to complete} The same as ``avg. document time to complete'' but only for documents with multiple collaborators.
     \bitem{Avg. non-collaborative document time to complete} The same as the ``avg. document time to complete'' but only for documents with no collaborators (one editor).
     \bitem{Completed documents} The number of documents completed in a week where the target worker is the primary editor.
     \bitem{Completed collaborative documents} The same as ``completed documents'' but only for documents with multiple collaborators.
     \bitem{Others' documents read/edited before completion} The number of documents read or edited in a week for which the target worker is not the primary editor.
     \bitem{Word out-of-hours session time} The subset of Word session time that occurred outside the individual's working day (see above).
 \end{itemize} 
 \subsubsection{Total out-of-hours session time} 
 The sum of Outlook, Teams, and Word out-of-hours session time, representing a maximum of the out-of-hours activity on these apps.

\subsection{Job Titles and Occupation Classification} \label{app:jobtitles}

Twenty-five of the sixty-six participating firms provided departments and/or job titles for the recruited workers. Where possible, we have used this information to impute occupation. When workers had missing job titles and departments that were not informative about work tasks (for example, a geographic region) we did not attempt to impute occupation. We classify workers into Standard Occupational Classification major groups by prompting GPT-4o (model version November 20, 2024) with unique industry and job title and/or department combinations across 5 different runs. We considered unique combinations of industry, department, and job title to be classifiable into a major occupational group if at least 3 or the 5 runs agreed. Combinations without majority agreement (accounting for 1\% of workers with classifiable information) were left unclassified, resulting in a total of 2,073 workers we can categorize into major groups. 

Table \ref{table:occupation_table} shows the distribution of workers across groups, and also reports the share of workers classified in each occupation group that had above-median work in the three applications we study during the pre-period. While the sample of workers with imputed occupation is too small for meaningful subgroup analyses, it confirms that the participating workers span a range of work responsibilities, but are generally white-collar knowledge workers.

\subsection{Coworkers \label{app:coworkers}}
\begin{figure}[ht!]
    \centering
    
    \includegraphics[width=0.5\linewidth]{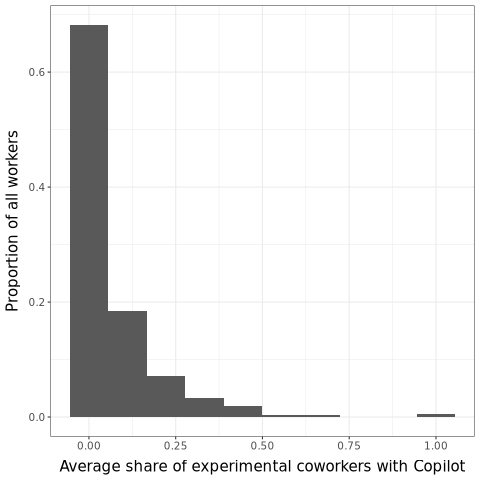}
    \caption{Distribution of share of coworkers with Copilot}
    \label{fig:coworker_share}
    \begin{tablenotes}
        This figure shows the distribution of the share of each worker's coworkers who were in the experiment and received Copilot, averaged across post-period months 4-6. 70\% of workers had on average (across weeks) less than 1 experimental coworker with Copilot, and the median worker had no experimental coworkers with Copilot. The average worker has 7 coworkers. 
    \end{tablenotes}
\end{figure}

We use telemetry data to identify the closest coworkers of each person in the experiment based on a weighted sum of their joint meetings, calls, emails, chats, and document collaborations. For each pair of workers, we calculate each metric and divide by the firm-specific average. We add these normalized numbers together for an interaction score and call someone a ``coworker'' if the score is greater than 8.0. (Since there are five metrics, if a pair of workers have an entirely average level of interaction, their score would be 5.) If an individual has more than 15 coworkers, we take only the 15 with the highest interaction score. The average worker has 7 coworkers. Our data use agreements prevent us from tracking individual work outcomes for the coworkers of our participating workers unless the coworkers are also study participants. However, we can aggregate across these workers to measure, for example, what share of an individual's coworkers have a Copilot license during the study period. Figure \ref{fig:coworker_share} shows the distribution of that share. 

\begin{table}[htbp]
    \caption{Occupational Classification and App Use}
    \resizebox{\textwidth}{!}{%
\begin{tabular}{>{\arraybackslash}p{8cm} >{\centering\arraybackslash}p{1.5cm} >{\centering\arraybackslash}p{1.3cm} >{\centering\arraybackslash}p{1.3cm} >{\centering\arraybackslash}p{1.3cm}}
\hline
& & \multicolumn{3}{c}{\textbf{Above-median use of}} \\
Group & Proportion & Outlook & Teams & Word \\
\hline
Management & 0.280 & 0.539 & 0.515 & 0.460 \\
Business and Financial Operations & 0.268 & 0.525 & 0.358 & 0.464 \\
Computer and Mathematical & 0.122 & 0.360 & 0.486 & 0.229 \\
Sales and Related & 0.110 & 0.581 & 0.240 & 0.358 \\
Architecture and Engineering & 0.068 & 0.357 & 0.429 & 0.514 \\

Office and Administrative Support & 0.053 & 0.486 & 0.294 & 0.284 \\
Legal & 0.018 & 0.789 & 0.105 & 0.500 \\
Arts, Design, Entertainment, Sports, and Media & 0.017 & 0.314 & 0.229 & 0.286 \\
Life, Physical, and Social Science & 0.010 & 0.095 & 0.095 & 0.048 \\
Other & 0.054 & 0.387 & 0.351 & 0.577 \\
\hline
\end{tabular}%
}

    \begin{tablenotes}
        This table reports the distribution of workers across occupation groups for the subset of workers for whom we were able to obtain meaningful job title and/or department information (2,073, or 29\%), as well as the proportion of each group who had above-median pre-period use of Outlook, Teams, and Word.
    \end{tablenotes}
    \label{table:occupation_table}
\end{table}

\begin{table}[htbp]
    \caption{Pre-Period Balance Test} 
    % \label{fig:balance}
    \resizebox{\textwidth}{!}{%
\begin{tabular}{>{\arraybackslash}p{9cm} >{\centering\arraybackslash}p{2.5cm} >{\centering\arraybackslash}p{2.5cm} > {\centering\arraybackslash}p{2.5cm}}
\hline
\textbf{Main Outcome Means} & \textbf{Control} & \textbf{Treated} & \textbf{$p$-value} \\
\hline
Outlook total session time & 11.73 & 11.58 & 0.38 \\
Outlook sessions & 31.91 & 31.44 & 0.094 \\
Email-free work time & 26.10 & 26.27 & 0.32 \\
Outlook out-of-hours session time & 2.22 & 2.17 & 0.29 \\
Unique email threads replied to & 14.41 & 14.60 & 0.57 \\
Time to reply from email delivery & 16.35 & 16.75 & 0.020 \\
Emails read & 161.37 & 160.32 & 0.73 \\
Teams meeting time & 5.17 & 5.28 & 0.26 \\
Total Teams meetings attended & 9.84 & 10.01 & 0.34 \\
Share of mtgs. attended early/late (\%) & 30.36 & 30.92 & 0.19 \\
Recurring Teams meeting time & 2.28 & 2.35 & 0.20 \\
Word total session time & 1.67 & 1.58 & 0.16 \\
Avg. document time to complete & 191.11 & 182.14 & 0.20 \\
Avg. collaborative document TTC & 276.04 & 299.36 & 0.14 \\
Avg. non-collaborative document TTC & 163.61 & 152.75 & 0.13 \\
Completed documents & 0.78 & 0.74 & 0.23 \\
Completed collaborative documents & 0.14 & 0.13 & 0.22 \\
Total out-of-hours session time & 2.76 & 2.69 & 0.20 \\
\\
\hline
\textbf{Additional Outcome Means} & \textbf{Control} & \textbf{Treated} & \textbf{$p$-value} \\
\hline 
Total emails in email threads replied to & 3.86 & 3.86 & 0.96 \\
Average email read duration (s) & 83.77 & 83.70 & 0.97 \\
Total recurring meetings & 4.06 & 4.19 & 0.15 \\
Big, short meetings & 1.08 & 1.09 & 0.86 \\
Big, long meetings & 0.89 & 0.90 & 0.70 \\
Small, short meetings & 6.63 & 6.83 & 0.11 \\
Small, long meetings & 1.19 & 1.23 & 0.13 \\
Teams out-of-hours meeting time & 0.37 & 0.35 & 0.077 \\
Word out-of-hours session time & 0.34 & 0.33 & 0.35 \\
Others' documents read/edited before completion & 0.41 & 0.39 & 0.43 \\
Number of email recipients (tens) & 4.09 & 4.49 & 0.027 \\
Number of unique people met 1-1 (total, tens) & 13.90 & 14.15 & 0.43 \\
\\
\hline
\textbf{Number of workers} & 3453 & 3684 & --- \\
\hline
\end{tabular}%
}
    \begin{tablenotes} This table shows average pre-period mean weekly outcomes for control and treatment group workers. Few significant differences were found from a \textit{t}-test on individual workers' pre-period means. All time metrics in hours. A worker is a primary editor if they make the most edits on a document. Time to complete (TTC) and completed document counts are calculated only for documents where the target worker is the primary editor.
    \end{tablenotes}
    
    \label{table:balance_table}
\end{table}

\FloatBarrier
\section{Adoption}
\setcounter{table}{0}
\setcounter{figure}{0}

\begin{table}[htb!]
    \caption{Copilot Usage, Overall and Across Apps} \label{tab:use_correl}
    \resizebox{\textwidth}{!}{%
\begin{tabular}{>{\arraybackslash}p{2.2cm} >{\centering\arraybackslash}p{2cm} >{\centering\arraybackslash}p{2cm} >{\centering\arraybackslash}p{2cm} >{\centering\arraybackslash}p{2cm} >{\centering\arraybackslash}p{2cm} >{\centering\arraybackslash}p{2.0cm}}
\hline
& \textbf{Teams} & \textbf{Word} & \textbf{Outlook} & \textbf{M365 Chat} & \textbf{PowerPoint} & \textbf{Excel} \\
\hline
Overall usage & 0.661 & 0.542 & 0.493 & 0.421 & 0.253 & 0.086 \\
\\[-6pt]
\hdashline\\[-4pt]
Teams & 1 & 0.807 & 0.799 & 0.918 & 0.849 & 0.896 \\
Word & 0.649 & 1 & 0.658 & 0.693 & 0.799 & 0.809 \\
Outlook & 0.608 & 0.625 & 1 & 0.643 & 0.667 & 0.741 \\
M365 Chat & 0.638 & 0.612 & 0.550 & 1 & 0.704 & 0.785 \\
PowerPoint & 0.360 & 0.457 & 0.342 & 0.422 & 1 & 0.568 \\
Excel & 0.127 & 0.152 & 0.129 & 0.160 & 0.193 & 1 \\
\hline
\end{tabular}%
}
    \begin{tablenotes} This table shows average weekly usage of Copilot through each integrated application among treated workers in post-period months 4-6. The first row reports overall usage rates. The following rows are conditional usage rates: given that a worker used the app Copilot in that column, what is the probability they also used the app Copilot in that row? Cells involving Teams Copilot only consider workers at Teams-using firms; cells involving Word Copilot only consider above-median Word users as in the main results.
    \end{tablenotes}
    % \label{table:usage_correlation}
\end{table}

\begin{table}[htbp]
    \caption{Adoption Regressions with Firm-Level Average Characteristics}\resizebox{\textwidth}{!}{%
\begingroup
\centering
\begin{tabular}{lccccccccc}
   \midrule \midrule
   Dependent Variable: & \multicolumn{9}{c}{Individual average Copilot use}\\
   Model:                                                       & (1)   & (2)   & (3)     & (4)     & (5)   & (6)   & (7)   & (8)     & (9)\\  
   \midrule
   % \emph{Variables}\\
   Firm average emails read (thousands)                         &       &       & -0.532  & -0.377  &       &       &       &         & -0.573\\   
                                                                &       &       & (0.177) & (0.210) &       &       &       &         & (0.251)\\   
   Firm average unique email threads replied to (thousands)     &       &       & 3.67    & 4.38    &       &       &       &         & 8.33\\   
                                                                &       &       & (1.63)  & (1.82)  &       &       &       &         & (2.31)\\   
   Firm average Teams meeting time (hours)                      &       &       & 0.021   & 0.006   &       &       &       &         & 0.005\\   
                                                                &       &       & (0.007) & (0.008) &       &       &       &         & (0.010)\\   
   Firm average total Teams meetings attended                   &       &       & 0.011   & 0.013   &       &       &       &         & 0.008\\   
                                                                &       &       & (0.004) & (0.004) &       &       &       &         & (0.005)\\   
   Firm average Word total session time (hours)                 &       &       & 0.042   & 0.041   &       &       &       &         & 0.042\\   
                                                                &       &       & (0.011) & (0.012) &       &       &       &         & (0.016)\\   
   Firm average others' documents read/edited before completion &       &       & 0.040   & 0.001   &       &       &       &         & -0.036\\   
                                                                &       &       & (0.032) & (0.035) &       &       &       &         & (0.043)\\   
   Firm average number of email recipients (tens)               &       &       & 0.090   & 0.098   &       &       &       &         & 0.113\\   
                                                                &       &       & (0.023) & (0.025) &       &       &       &         & (0.031)\\   
   Firm average number of unique people met 1-1 (total, tens)   &       &       & -0.007  & -0.015  &       &       &       &         & -0.011\\   
                                                                &       &       & (0.015) & (0.017) &       &       &       &         & (0.020)\\   
   Experiment start calendar month: 2023-10-01                  &       &       &         &         &       &       &       & 0.006   & -0.031\\   
                                                                &       &       &         &         &       &       &       & (0.027) & (0.030)\\   
   Experiment start calendar month: 2023-11-01                  &       &       &         &         &       &       &       & -0.092  & -0.070\\   
                                                                &       &       &         &         &       &       &       & (0.026) & (0.029)\\   
   Experiment start calendar month: 2023-12-01                  &       &       &         &         &       &       &       & -0.020  & -0.024\\   
                                                                &       &       &         &         &       &       &       & (0.029) & (0.034)\\   
   Experiment start calendar month: 2024-01-01                  &       &       &         &         &       &       &       & 0.161   & 0.087\\   
                                                                &       &       &         &         &       &       &       & (0.028) & (0.034)\\   
   Experiment start calendar month: 2024-02-01                  &       &       &         &         &       &       &       & -0.016  & 0.017\\   
                                                                &       &       &         &         &       &       &       & (0.038) & (0.042)\\   
   Experiment start calendar month: 2024-03-01                  &       &       &         &         &       &       &       & -0.084  & -0.044\\   
                                                                &       &       &         &         &       &       &       & (0.031) & (0.036)\\   
   Experiment start calendar month: 2024-04-01                  &       &       &         &         &       &       &       & 0.022   & -0.034\\   
                                                                &       &       &         &         &       &       &       & (0.036) & (0.040)\\   
   Other controls                                               & Yes   & Yes   & Yes     & Yes     & Yes   & Yes   & Yes   &         & Yes\\  
   \midrule
   \emph{Fixed-effects}\\
   Firm                                                         &       &       &         &         & Yes   & Yes   &       &         & \\  
   Industry                                                     &       &       &         &         &       &       & Yes   & Yes     & Yes\\  
   \midrule
   % \emph{Fit statistics}\\
   Observations                                                 & 3,422 & 3,422 & 3,684   & 3,422   & 3,684 & 3,422 & 3,684 & 3,684   & 3,422\\  
   R$^2$                                                        & 0.092 & 0.092 & 0.087   & 0.120   & 0.181 & 0.213 & 0.029 & 0.081   & 0.151\\  
   \midrule \midrule
   %                                                              &       &       &         &         &       &       &       &         &  \tabularnewline 
   % \\
\end{tabular}
\par\endgroup
}

    \begin{tablenotes}
        This table reports the coefficients on firm pre-period behavior and experiment start month from the regressions reported in Table \ref{table:horserace}. Coefficients on experiment start months are relative to a September 2023 baseline.
    \end{tablenotes}
    \label{table:horserace_firm_start_month_coefs}
\end{table}

\FloatBarrier

\clearpage
\setcounter{table}{0}
\setcounter{figure}{0}
\section{Additional Outcomes}

\begin{table}[htbp]
    \caption{Effects of Copilot in All Post-Period Weeks}
    \resizebox{\textwidth}{!}{%
\begin{tabular}{>{\arraybackslash}p{6cm} >{\centering\arraybackslash}p{2cm} >{\centering\arraybackslash}p{2cm} >{\centering\arraybackslash}p{2cm} >{\centering\arraybackslash}p{1.3cm} >{\centering\arraybackslash}p{1.3cm}}
\hline
& (1) & (2) & (3) & (4) & (5) \\
\textbf{Metric} & \textbf{Pre-period mean (SD)} &  \textbf{All-months OLS (SE)} & \textbf{All-months IV (SE)} & \textbf{Workers} & $\mathbf{n}$ \\
\hline
Outlook total session time & 11.65 & -1.29 & -1.91 & 6441 & 313461 \\
& (6.81) & (0.10) & (0.15) & & \\
Outlook sessions & 31.67 & -3.58 & -5.30 & 6441 & 313461 \\
& (11.08) & (0.24) & (0.36) & & \\
Email-free work time & 26.18 & 1.44 & 2.13 & 6441 & 313461 \\
& (6.77) & (0.11) & (0.16) & & \\
Outlook out-of-hours session time & 2.19 & -0.30 & -0.45 & 6441 & 313461 \\
& (1.84) & (0.031) & (0.046) & & \\
Emails read & 160.83 & -7.46 & -11.08 & 6406 & 280932 \\
& (117.74) & (1.28) & (1.91) & & \\
Unique email threads replied to & 14.51 & -0.18 & -0.26 & 6441 & 337149 \\
& (14.17) & (0.14) & (0.20) & & \\
Time to reply from email delivery & 16.55 & -0.17 & -0.25 & 6435 & 297804 \\
& (5.34) & (0.13) & (0.19) & & \\
\\[-6pt]
\hdashline\\[-4pt]
Teams meeting time & 5.22 & 0.081 & 0.11 & 6170 & 221116 \\
& (3.65) & (0.046) & (0.064) & & \\
Total Teams meetings attended & 9.93 & 0.20 & 0.28 & 6170 & 319836 \\
& (6.93) & (0.093) & (0.13) & & \\
Share of mtgs. attended early/late (\%) & 30.65 & 0.52 & 0.71 & 6101 & 186486 \\
& (11.94) & (0.23) & (0.32) & & \\
Recurring Teams meeting time & 2.32 & 0.025 & 0.035 & 6170 & 268353 \\
& (1.93) & (0.025) & (0.034) & & \\
\\[-6pt]
\hdashline\\[-4pt]
Word total session time & 1.63 & 0.11 & 0.15 & 2525 & 132213 \\
& (1.58) & (0.055) & (0.075) & & \\
Avg. document time to complete & 186.54 & -4.93 & -6.57 & 2525 & 40366 \\
& (94.20) & (7.16) & (9.53) & & \\
Avg. collaborative document TTC & 287.85 & -41.60 & -52.72 & 1910 & 11836 \\
& (127.78) & (17.38) & (22.04) & & \\
Avg. non-collaborative document TTC & 158.07 & 0.55 & 0.73 & 2514 & 34383 \\
& (89.13) & (7.21) & (9.64) & & \\
Completed documents & 0.76 & 0.030 & 0.041 & 2525 & 104438 \\
& (0.80) & (0.021) & (0.029) & & \\
Completed collaborative documents & 0.14 & 0.012 & 0.016 & 2525 & 104438 \\
& (0.21) & (0.0062) & (0.0084) & & \\
\\[-6pt]
\hdashline\\[-4pt]
Total out-of-hours session time & 2.72 & -0.24 & -0.35 & 6304 & 307005 \\
& (2.16) & (0.042) & (0.061) & & \\
\hline
\end{tabular}%
}
    \begin{tablenotes} This table repeats main text Table \ref{table:main_results} using data from every post-period week instead of only months 4-6. For easier comparison with the main specification, the IV estimates still estimate the effect of any use of Copilot in post-period months 4-6, instrumented with treatment assignment.
    \end{tablenotes}
    \label{table:all_months_iv}
\end{table}

\begin{table}[htbp]
  \caption{Effects of Copilot on Additional Outcome Measures} 
    \label{table:app_results}    \resizebox{\textwidth}{!}{%
\begin{tabular}{>{\arraybackslash}p{6cm} >{\centering\arraybackslash}p{2cm} >{\centering\arraybackslash}p{1.3cm} >{\centering\arraybackslash}p{1.3cm} >{\centering\arraybackslash}p{1.3cm} >{\centering\arraybackslash}p{1.3cm}}
\hline
\textbf{Metric} & \textbf{Pre-period mean (SD)} & \textbf{OLS (SE)} & \textbf{IV (SE)} & \textbf{Workers} & $\mathbf{n}$ \\
\hline
Total emails in email threads replied to & 3.86 & -0.019 & -0.027 & 6432 & 235121 \\
& (0.84) & (0.018) & (0.026) & & \\
Average email read duration (s) & 83.73 & -1.82 & -2.72 & 6387 & 223680 \\
& (64.28) & (1.26) & (1.89) & & \\
\\[-6pt]
\hdashline\\[-4pt]
Total recurring meetings & 4.13 & 0.11 & 0.16 & 6170 & 202314 \\
& (3.45) & (0.051) & (0.071) & & \\
Big, short meetings & 1.08 & 0.061* & 0.085* & 6170 & 202314 \\
& (1.26) & (0.021) & (0.030) & & \\
Big, long meetings & 0.90 & 0.024 & 0.034 & 6170 & 202314 \\
& (0.78) & (0.013) & (0.019) & & \\
Small, short meetings & 6.73 & 0.13 & 0.18 & 6170 & 202314 \\
& (4.63) & (0.070) & (0.098) & & \\
Small, long meetings & 1.21 & 0.0011 & 0.0015 & 6170 & 202314 \\
& (0.98) & (0.018) & (0.024) & & \\
Teams out-of-hours meeting time & 0.36 & 0.0087 & 0.012 & 5850 & 239620 \\
& (0.50) & (0.019) & (0.027) & & \\
\\[-6pt]
\hdashline\\[-4pt]
Word out-of-hours session time & 0.33 & 0.037 & 0.051 & 2459 & 101985 \\
& (0.39) & (0.025) & (0.034) & & \\
\hline
\multicolumn{6}{l}{\footnotesize Significance: ** $q < 0.01$, * $q < 0.05$} \\
\end{tabular}%
}

    \begin{tablenotes} This table repeats the regressions in Table \ref{table:main_results} for some additional outcomes. Asterisks are computed from sharpened $q$-values that account for the total number of hypotheses tested in the main text and this appendix table. See the notes to Table \ref{table:main_results} for additional details.
    %All time metrics are in hours. 
    % Share of meetings attended late/early refers to the share of Teams meetings that a worker joined more than 5 minutes after the meeting started or left more than 5 minutes before it ended. We assign documents to a worker if they were the `primary editor', making the more most edits to the document. 
    %A worker is a primary editor if they make the most edits on a document. Time to complete (TTC) and completed document counts are calculated only for documents where the target worker is the primary editor.
    \end{tablenotes}
\end{table}

\begin{table}[htbp]
    \caption{Reduced Form Regressions of Outcomes on Application-Specific Copilot Usage}
    \resizebox{\textwidth}{!}{%
\begin{tabular}{>{\arraybackslash}p{7cm} >{\centering\arraybackslash}p{2.5cm} >{\centering\arraybackslash}p{2.5cm} >{\centering\arraybackslash}p{2.5cm} >{\centering\arraybackslash}p{2.5cm}}
\hline
& (1) & (2) & (3) & (4) \\
\textbf{Metric} & \textbf{Outlook Copilot} & \textbf{Teams Copilot} & \textbf{Word Copilot} & \textbf{Other Copilot} \\
\hline
Outlook total session time & -1.88 & -0.33 & -0.18 & -0.16 \\
& (0.19) & (0.20) & (0.19) & (0.22) \\
Outlook sessions & -6.07 & -0.35 & 0.27 & -0.56 \\
& (0.45) & (0.45) & (0.45) & (0.50) \\
Email-free work time & 2.16 & 0.25 & 0.17 & 0.26 \\
& (0.21) & (0.22) & (0.21) & (0.23) \\
Outlook out-of-hours session time & -0.45 & -0.091 & -0.029 & 0.0041 \\
& (0.053) & (0.057) & (0.055) & (0.061) \\
Emails read & -11.86 & -1.85 & -0.84 & -0.88 \\
& (2.35) & (2.55) & (2.31) & (2.67) \\
Unique email threads replied to & 0.31 & 0.16 & 0.026 & -0.14 \\
& (0.22) & (0.24) & (0.23) & (0.25) \\
Time to reply from email delivery & -0.0061 & -0.56 & -0.38 & 0.23 \\
& (0.22) & (0.26) & (0.24) & (0.26) \\
\\[-6pt]
\hdashline\\[-4pt]
Teams meeting time & 0.038 & 0.061 & 0.14 & 0.10 \\
& (0.076) & (0.086) & (0.080) & (0.088) \\
Total Teams meetings attended & 0.17 & 0.43 & 0.18 & 0.25 \\
& (0.14) & (0.16) & (0.15) & (0.17) \\
Share of mtgs. attended early/late (\%) & 0.23 & 0.010 & 0.39 & -0.13 \\
& (0.37) & (0.42) & (0.37) & (0.42) \\
Recurring Teams meeting time & 0.019 & 0.086 & 0.0023 & -0.022 \\
& (0.038) & (0.044) & (0.040) & (0.047) \\
\\[-6pt]
\hdashline\\[-4pt]
Word total session time & -0.067 & 0.10 & 0.36 & -0.033 \\
& (0.090) & (0.10) & (0.10) & (0.10) \\
Avg. document time to complete & -19.19 & 9.03 & -21.03 & 12.41 \\
& (11.64) & (13.26) & (12.85) & (13.37) \\
Avg. collaborative document TTC & -56.64 & -11.36 & -29.57 & 34.57 \\
& (31.26) & (33.60) & (32.87) & (33.22) \\
Avg. non-collaborative document TTC & -1.04 & 14.34 & -29.08 & 7.58 \\
& (11.84) & (13.60) & (13.99) & (14.14) \\
Completed documents & -0.063 & 0.049 & 0.12 & 0.012 \\
& (0.038) & (0.039) & (0.037) & (0.040) \\
Completed collaborative documents & -0.024 & 0.0087 & 0.023 & 0.016 \\
& (0.011) & (0.011) & (0.010) & (0.011) \\
\\[-6pt]
\hdashline\\[-4pt]
Total out-of-hours session time & -0.37 & 0.0011 & 0.28 & -0.13 \\
& (0.070) & (0.077) & (0.076) & (0.081) \\
\hline
\end{tabular}%
}
 \label{tab:indiv_copilot_redform}
    \begin{tablenotes} This table shows estimates resulting from reduced-form regressions of our main outcomes on weekly indicators for any use of Copilot through each of our studied applications (pooling the remaining applications) These results include all fixed effect controls described in Equation \ref{regression}. Note that the treatment in these regressions is endogenous (Copilot use, not randomized access) and we cannot estimate a LATE because we have only one instrument (treatment assignment) for multiple endogenous regressors. We still control for some confounding variation through the difference-in-difference framework, but nonetheless interpret these estimates as suggestive correlations rather than causal effects.
    \end{tablenotes}
    % \label{table:reduced_form}
\end{table}

\begin{table}[htbp]
    \caption{Treatment Effect Heterogeneity by Pre-Period Behavior}
    \label{table:heterogeneity}
    \subcaption{Best Linear Predictors of Conditional Average Treatment Effects (CATEs) by Proxy CATEs}
    \centering
\begin{tabular}{lcc|cc}

\hline
& \multicolumn{2}{c}{Elastic Net} & \multicolumn{2}{c}{Neural Network} \\

\cline{2-3} \cline{4-5}
& ATE ($\beta_1$) & HET ($\beta_2$) & ATE ($\beta_1$) & HET ($\beta_2$) \\

\hline

Outlook total session time & -1.364 & 0.282 & -1.339 & 0.043 \\
& (-1.884, -0.846) & (-0.518, 1.110) & (-1.858, -0.821) & (-0.136, 0.229) \\
& [0.000] & [0.440] & [0.000] & [0.626] \\

\hline

\end{tabular}

    \vspace{0.5em}
    \subcaption{Group-Average Treatment Effects by Proxy CATEs}
    \resizebox{\textwidth}{!}{%
\centering

\begin{tabular}{lccc|ccc}

\hline
& \multicolumn{3}{c}{Elastic Net} & \multicolumn{3}{c}{Nnet} \\

\cline{2-4} \cline{5-7}
& 20\% Most & 20\% Least & Difference & 20\% Most ($G_5$) & 20\% Least ($G_1$) & Difference \\

\hline

Outlook total session time  
& -1.649 & -1.177 & 0.532 & -1.467 & -1.029 & 0.466 \\
& (-3.090, -0.251) & (-2.24, -0.139) & (-1.187, 2.255) & (-2.741, -0.213) & (-2.165, 0.106) & (-1.279, 2.197) \\
& [0.021] & [0.027] & [0.551] & [0.022] & [0.075] & [0.505] \\

\hline

\end{tabular}% 
}

    \begin{tablenotes} 
    These tables show estimates of heterogeneous effects of Copilot on Outlook total session time using predictive learners from \citet{chernozhukov2025heteffects}. Elastic net and neural net methods were used to create proxy treatment effect estimates conditional on pre-period behaviors from Table \ref{table:horserace}. The first panel shows estimates of the best linear predictor of the true conditional average treatment effects (CATEs) by the proxy estimates. The second panel shows estimates of the true CATEs, grouping workers by quintile of their proxy estimate. All statistics shown are median estimates among 30 sample splits.
    \end{tablenotes}
\end{table}

\begin{table}[htbp]
    \caption{Effects of Copilot - Split by Coworkers with Copilot}
    \label{table:split_results_teammates}    \resizebox{\textwidth}{!}{%
\begin{tabular}{>{\arraybackslash}p{6.1cm} >{\centering\arraybackslash}p{1.2cm} >{\centering\arraybackslash}p{1.2cm} >{\centering\arraybackslash}p{1.3cm} >{\centering\arraybackslash}p{1.3cm} >{\centering\arraybackslash}p{1.3cm} >{\centering\arraybackslash}p{1.3cm} >{\centering\arraybackslash}p{1.2cm} >{\centering\arraybackslash}p{1.2cm}}
\hline
\textbf{Outcome} & \multicolumn{2}{c}{\textbf{Pre-period }} & \multicolumn{2}{c}{\textbf{OLS (SE)}} & \multicolumn{2}{c}{\textbf{IV (SE)}} & \multicolumn{2}{c}{$\mathbf{n}$} \\
&\multicolumn{2}{c}{\textbf{mean (SD)}} \\
& (1) & (2) & (3) & (4) & (5) & (6) & (7) & (8) \\
& Below & Above & Below & $\Delta(\text{Above})$ & Below & $\Delta(\text{Above})$ & Below & Above \\
\hline
Outlook total session time & 11.40 & 11.88 & -1.30** & -0.17 & -1.70** & -0.69* & 115641 & 134504 \\
& (7.22) & (6.42) & (0.19) & (0.26) & (0.24) & (0.26) & & \\
Outlook sessions & 30.65 & 32.57 & -3.25** & -0.68 & -4.42** & -1.82* & 115641 & 134504 \\
& (12.13) & (10.03) & (0.40) & (0.59) & (0.52) & (0.61) & & \\
Email-free work time & 26.61 & 25.81 & 1.34** & 0.32 & 1.77** & 0.88* & 115641 & 134504 \\
& (7.28) & (6.26) & (0.20) & (0.28) & (0.25) & (0.28) & & \\
Outlook out-of-hours session time & 2.17 & 2.22 & -0.33** & -0.0062 & -0.42** & -0.13 & 115641 & 134504 \\
& (1.91) & (1.77) & (0.058) & (0.077) & (0.070) & (0.074) & & \\
Emails read & 163.63 & 158.39 & -6.13 & -1.50 & -7.62* & -5.03 & 102415 & 121287 \\
& (129.55) & (107.73) & (2.50) & (3.27) & (2.93) & (3.07) & & \\
Unique email threads replied to & 14.09 & 14.87 & -0.020 & -0.24 & 0.16 & -0.79* & 123711 & 144176 \\
& (14.26) & (14.03) & (0.25) & (0.34) & (0.29) & (0.31) & & \\
Time to reply from email delivery & 16.31 & 16.77 & -0.43 & 0.20 & -0.41 & -0.12 & 105865 & 129256 \\
& (5.44) & (5.26) & (0.25) & (0.34) & (0.29) & (0.30) & & \\
\\[-6pt]
\hdashline\\[-4pt]
Teams meeting time & 4.64 & 5.77 & 0.18 & -0.14 & 0.27* & -0.25* & 73069 & 81934 \\
& (3.71) & (3.52) & (0.077) & (0.11) & (0.090) & (0.096) & & \\
Total Teams meetings attended & 8.82 & 10.96 & 0.37 & -0.28 & 0.81** & -0.97** & 120781 & 132942 \\
& (7.18) & (6.55) & (0.16) & (0.22) & (0.18) & (0.19) & & \\
Share of mtgs. attended early/late (\%) & 32.57 & 28.90 & 0.062 & -0.18 & 0.13 & -0.19 & 58198 & 70462 \\
& (13.30) & (10.70) & (0.43) & (0.55) & (0.45) & (0.45) & & \\
Recurring Teams meeting time & 2.12 & 2.51 & 0.092 & -0.11 & 0.13* & -0.15* & 93202 & 109112 \\
& (2.01) & (1.87) & (0.041) & (0.058) & (0.047) & (0.050) & & \\
\\[-6pt]
\hdashline\\[-4pt]
Word total session time & 1.66 & 1.60 & 0.15 & -0.015 & 0.23 & -0.067 & 47668 & 57245 \\
& (1.62) & (1.54) & (0.099) & (0.13) & (0.11) & (0.11) & & \\
Avg. document time to complete & 181.22 & 191.12 & -8.87 & 10.49 & -4.03 & -2.42 & 12990 & 16693 \\
& (85.21) & (99.39) & (13.19) & (18.00) & (13.43) & (14.55) & & \\
Avg. collaborative document TTC & 278.47 & 294.27 & -10.54 & -62.68 & -32.10 & -48.48 & 3349 & 5351 \\
& (105.44) & (138.76) & (34.91) & (45.41) & (33.41) & (36.35) & & \\
Avg. non-collaborative document TTC & 157.59 & 158.48 & -10.47 & 22.07 & -2.45 & 5.40 & 11297 & 13955 \\
& (91.69) & (87.37) & (13.27) & (18.31) & (13.46) & (15.01) & & \\
Completed documents & 0.75 & 0.77 & 0.076 & -0.096 & 0.050 & -0.021 & 34743 & 42395 \\
& (0.83) & (0.78) & (0.037) & (0.051) & (0.042) & (0.046) & & \\
Completed collaborative documents & 0.12 & 0.15 & 0.028 & -0.032 & 0.016 & -0.0028 & 34743 & 42395 \\
& (0.21) & (0.21) & (0.010) & (0.014) & (0.012) & (0.013) & & \\
\\[-6pt]
\hdashline\\[-4pt]
Total out-of-hours session time & 2.66 & 2.78 & -0.25* & 0.0029 & -0.38** & 0.042 & 112474 & 132470 \\
& (2.22) & (2.11) & (0.077) & (0.10) & (0.091) & (0.097) & & \\
\hline
\multicolumn{9}{l}{\footnotesize Significance: ** $q < 0.01$, * $q < 0.05$} \\
\end{tabular}%
}

    \begin{tablenotes} This table replicates the analysis in Table \ref{table:main_results}, comparing estimated effects for a baseline group of workers who had a below-median number of close coworkers with Copilot, averaged over the post-period, to estimates for the corresponding above-median group. Significant values in Columns 4 and 6 therefore indicate significant differences in estimated effects. Standard errors are clustered by individual and $q$-values are calculated using adjustments for all hypotheses tested. 
    \end{tablenotes}
\end{table}

\begin{figure}[htb]
%\label{fig:app_event_studies}
    \centering
    \begin{subfigure}{0.31\textwidth}
        \centering
        
        \includegraphics[width=\textwidth]{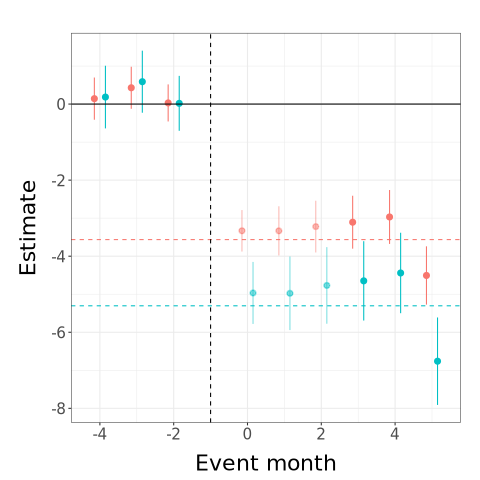}
        \caption{Outlook Sessions}
    \end{subfigure}
    \begin{subfigure}{0.31\textwidth}
        \centering
        
        \includegraphics[width=\textwidth]{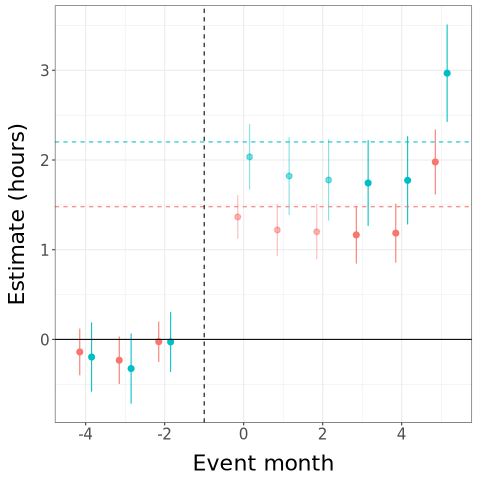}
        \caption{Email-Free Work Time       %\label{fig:event_concentoutlook}
        }
    \end{subfigure}
    \begin{subfigure}{0.31\textwidth}
        \centering
        
        \includegraphics[width=\textwidth]{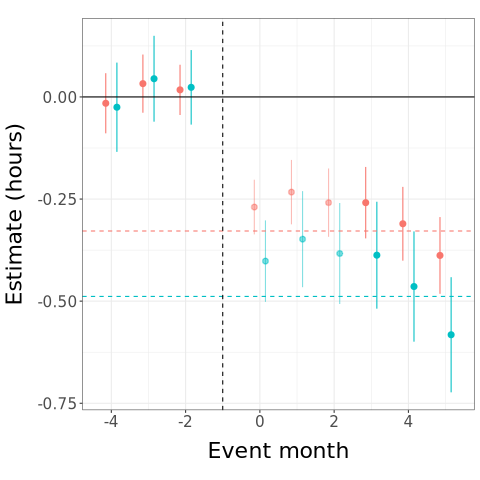}
        \caption{Outlook OOH Time %- Across Firms
       % \label{fig:event_word}
        }
    \end{subfigure}
    \begin{subfigure}{0.31\textwidth}
        \centering
        
        \includegraphics[width=\textwidth]{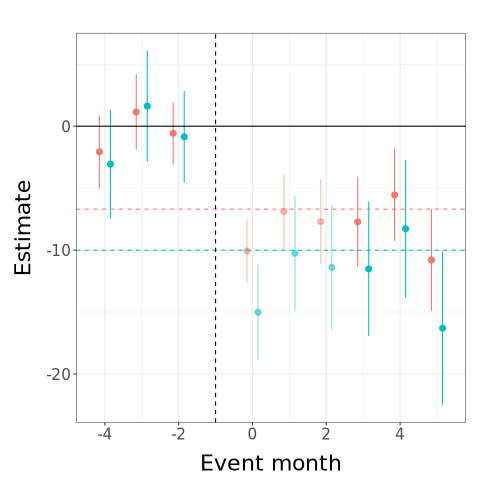}
        \caption{Emails Read %- Across Firms
       % \label{fig:event_word}
        }
    \end{subfigure}
    \begin{subfigure}{0.31\textwidth}
        \centering
         
        \includegraphics[width=\textwidth]{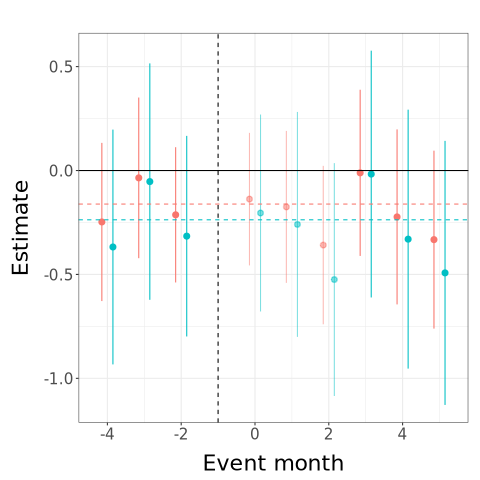}
       \caption{Email Threads Replied To
       % \label{fig:event_meetings}
       }
    \end{subfigure}
    \begin{subfigure}{0.31\textwidth}
        \centering
        
        \includegraphics[width=\textwidth]{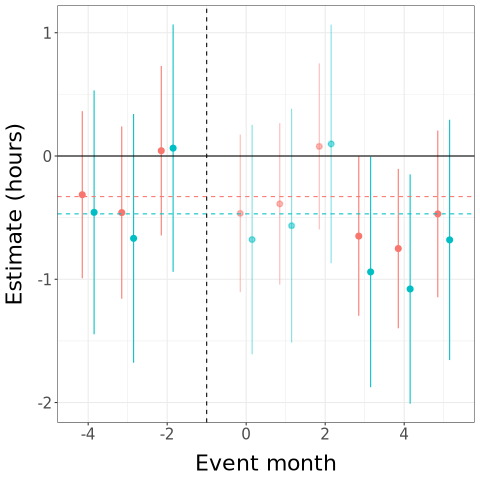}
        \caption{Time to Reply          %\label{fig:event_concentoutlook}
        }
    \end{subfigure}\\
    \begin{subfigure}{0.31\textwidth}
        \centering
        
        \includegraphics[width=\textwidth]{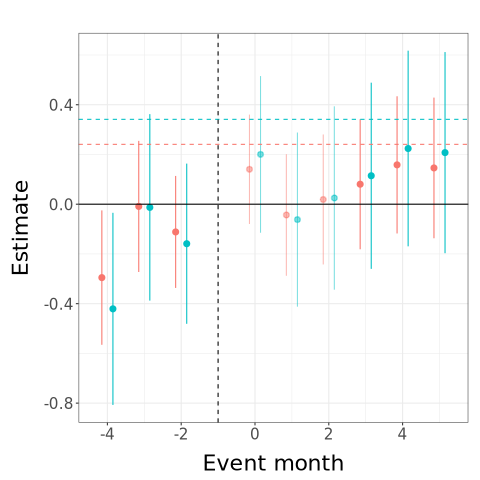}
        \caption{Teams Meetings        %\label{fig:event_concentoutlook}
        }
    \end{subfigure}
        \begin{subfigure}{0.31\textwidth}
        \centering
        
        \includegraphics[width=\textwidth]{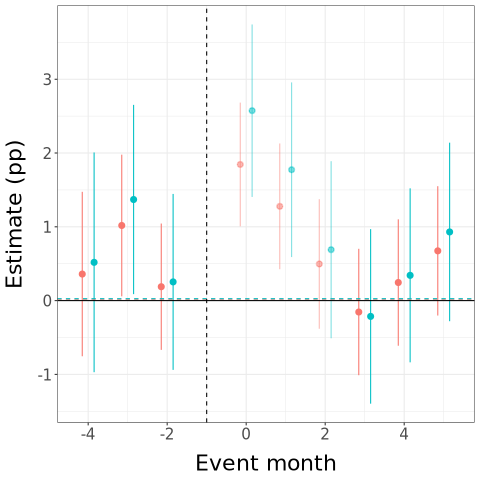}
        \caption{\% Meetings  Early/Late %- Across Firms
       % \label{fig:event_word}
        }
    \end{subfigure}
    \begin{subfigure}{0.31\textwidth}
        \centering
         
        \includegraphics[width=\textwidth]{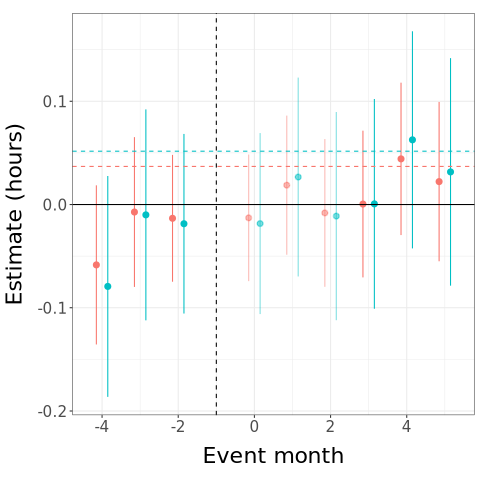}
       \caption{Recurring Meeting Time
       % \label{fig:event_meetings}
       }
    \end{subfigure}
    \begin{subfigure}{0.4\textwidth}
        \centering
         
        \includegraphics[width=0.5\textwidth]{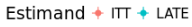}
    \end{subfigure}\\
    \caption{Event Study Plots for the Effect of Copilot on Other Outcomes\label{fig:event_studies_appendix}}
    \begin{tablenotes}
        These plots show the coefficients from regressions similar to Equation \eqref{regression} for the ITT (intent to treat) estimates and Equation \eqref{eqIV} for the LATE estimates. Vertical lines show 95\% confidence intervals based on individual-clustered standard errors. Horizontal dashed lines give the average estimated effect from Table \ref{table:main_results}. We include the coefficients for event months 0-2 for context.
    \end{tablenotes}
\end{figure}
\begin{figure}[htb]
%\label{fig:app_event_studies}
    \centering
        \begin{subfigure}{0.31\textwidth}
        \centering
        
        \includegraphics[width=\textwidth]{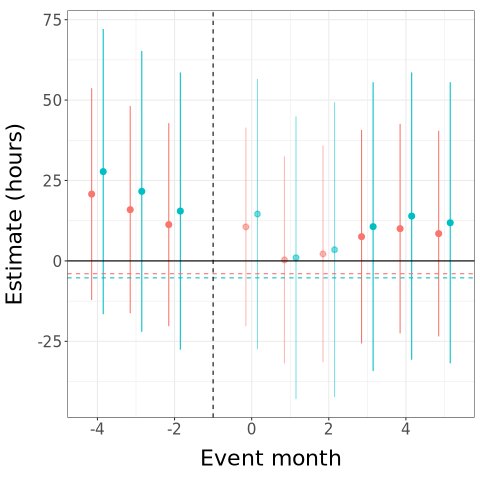}
        \caption{Document TTC
        %\label{fig:event_concentoutlook}
        }
    \end{subfigure}
        \begin{subfigure}{0.31\textwidth}
        \centering
        
        \includegraphics[width=\textwidth]{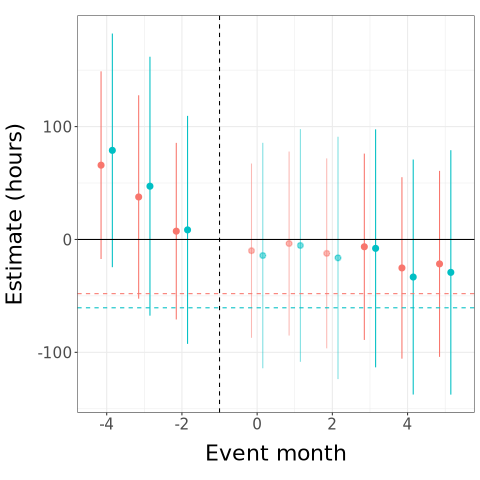}
        \caption{Collaborative Doc. TTC %- Across Firms
       % \label{fig:event_word}
        }
    \end{subfigure}
    \begin{subfigure}{0.31\textwidth}
        \centering
         
        \includegraphics[width=\textwidth]{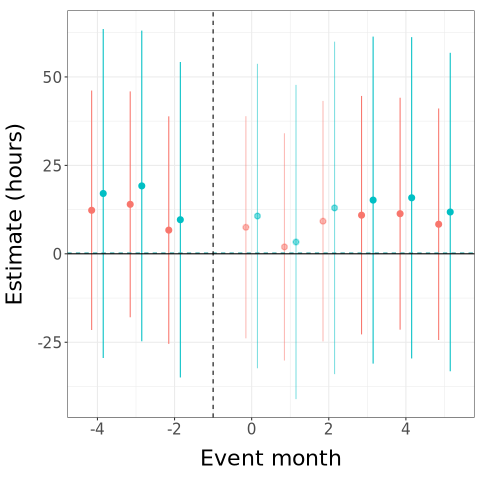}
       \caption{Non-Collab. Doc. TTC
       % \label{fig:event_meetings}
       }
    \end{subfigure}    
    \begin{subfigure}{0.31\textwidth}
        \centering
        
        \includegraphics[width=\textwidth]{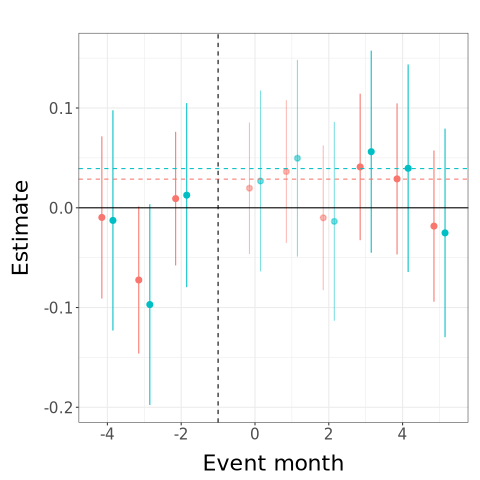}
        \caption{Completed Documents
        %\label{fig:event_concentoutlook}
        }
    \end{subfigure}
        \begin{subfigure}{0.31\textwidth}
        \centering
        
        \includegraphics[width=\textwidth]{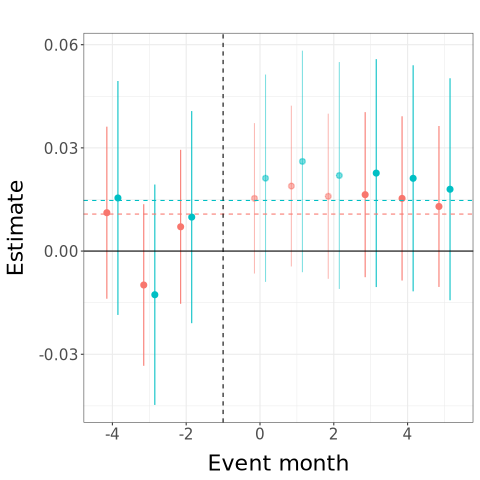}
        \caption{Completed Collab. Doc. %- Across Firms
       % \label{fig:event_word}
        }
    \end{subfigure}
    \begin{subfigure}{0.31\textwidth}
        \centering
        
        \includegraphics[width=\textwidth]{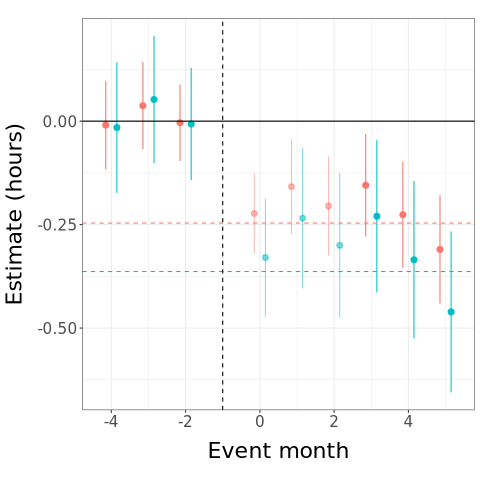}
        \caption{Total OOH Session Time %- Across Firms
       % \label{fig:event_word}
                }
    \end{subfigure}

    \caption{Event Study Plots for the Effect of Copilot on Other Outcomes - Continued\label{fig:event_studies_appendix_cont}}
    \begin{tablenotes}
        These plots show the coefficients from regressions similar to Equation \eqref{regression} for the ITT (intent to treat) estimates and Equation \eqref{eqIV} for the LATE estimates. Vertical lines show 95\% confidence intervals based on individual-clustered standard errors. Horizontal dashed lines give the average estimated effect from Table \ref{table:main_results}. We include the coefficients for event months 0-2 for context.
    \end{tablenotes}
\end{figure}

\begin{figure}[htbp]
%\label{fig:app_event_studies}
    \centering
    \begin{subfigure}{0.31\textwidth}
        \centering
        
        \includegraphics[width=\textwidth]{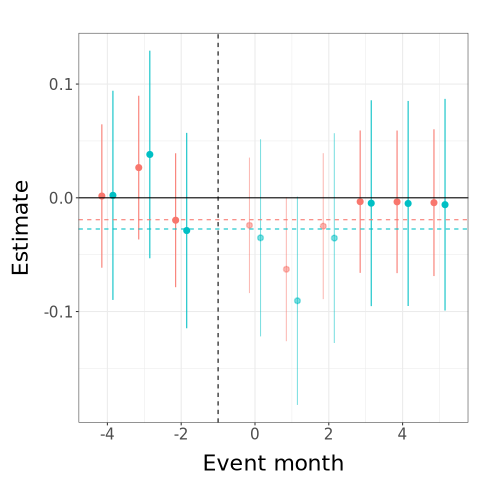}
        \caption{Emails in Threads Replied To %- Across Firms
       % \label{fig:event_word}
        }
    \end{subfigure}
    \begin{subfigure}{0.31\textwidth}
        \centering
        
        \includegraphics[width=\textwidth]{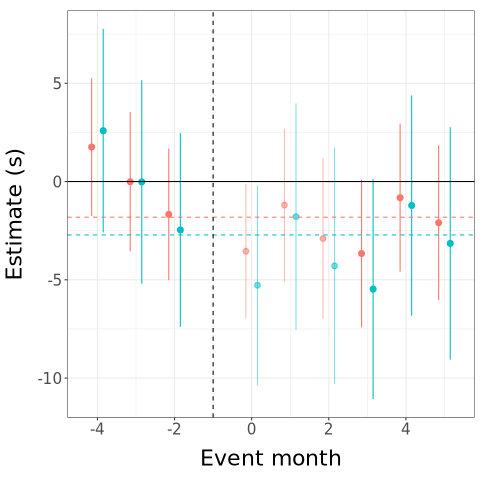}
        \caption{Avg. Email Read Duration %- Across Firms
       % \label{fig:event_word}
        }
    \end{subfigure}
    \begin{subfigure}{0.31\textwidth}
        \centering
        
        \includegraphics[width=\textwidth]{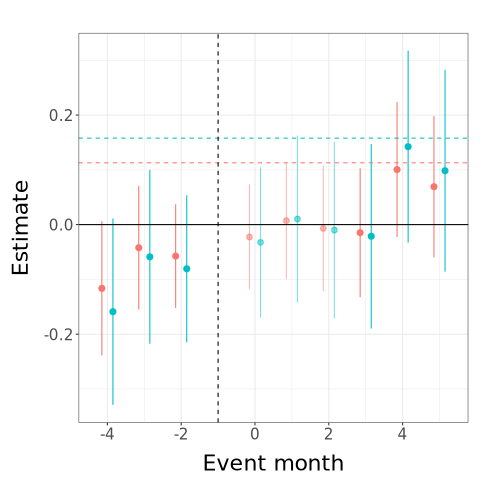}
        \caption{Recurring Meetings %- Across Firms
       % \label{fig:event_word}
        }
    \end{subfigure}
    \begin{subfigure}{0.31\textwidth}
        \centering
        
        \includegraphics[width=\textwidth]{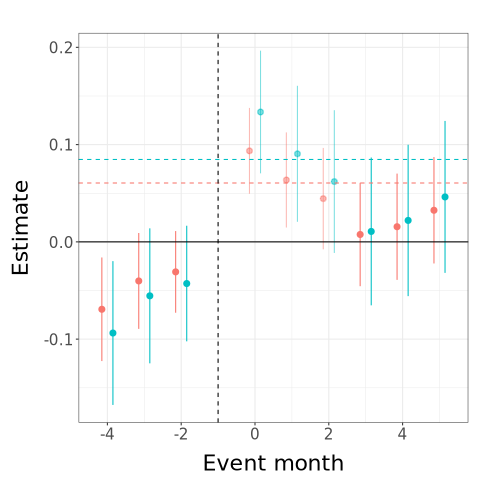}
        \caption{Big Short Meetings %- Across Firms
       % \label{fig:event_word}
        }
    \end{subfigure}
    \begin{subfigure}{0.31\textwidth}
        \centering
         
        \includegraphics[width=\textwidth]{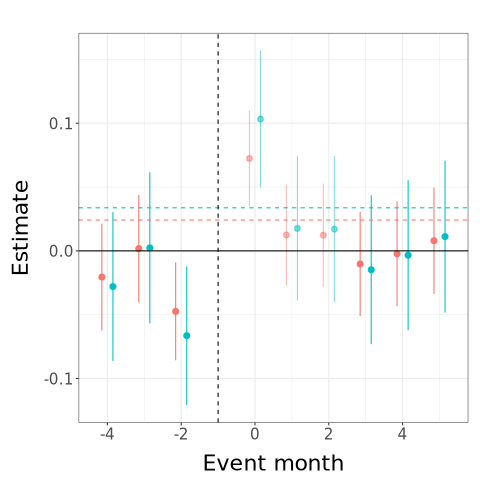}
       \caption{Big Long Meetings
       % \label{fig:event_meetings}
       }
    \end{subfigure}
    \begin{subfigure}{0.31\textwidth}
        \centering
        
        \includegraphics[width=\textwidth]{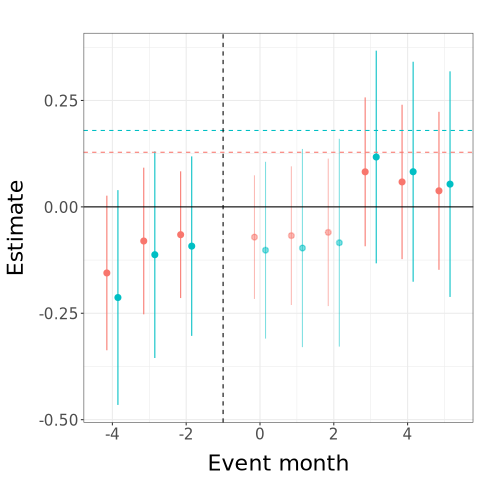}
        \caption{Small Short Meetings % Across Firms
       % \label{fig:event_word}
        }
    \end{subfigure}
    \begin{subfigure}{0.31\textwidth}
        \centering
         
        \includegraphics[width=\textwidth]{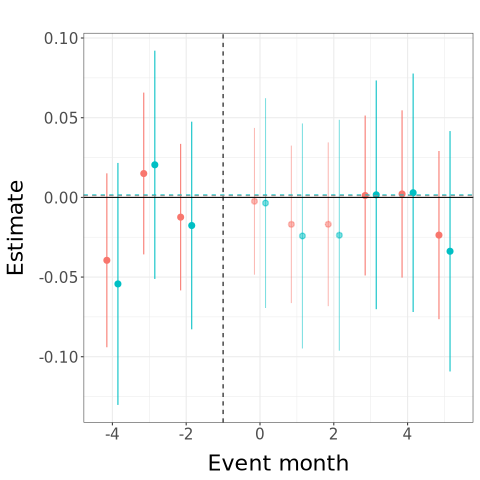}
       \caption{Small Long Meetings 
       % \label{fig:event_meetings}
       }
    \end{subfigure}
     \begin{subfigure}{0.31\textwidth}
        \centering
         
        \includegraphics[width=\textwidth]{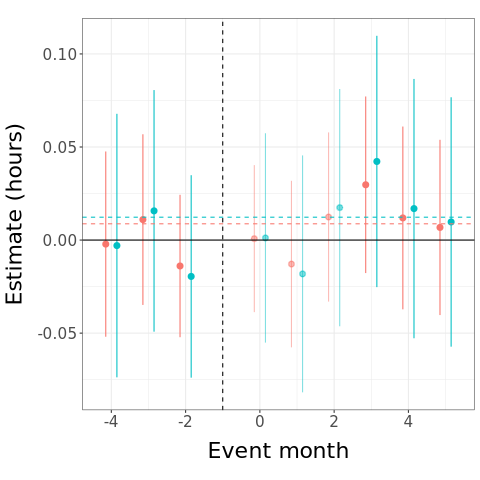}
       \caption{Teams OOH Meeting Time
       % \label{fig:event_meetings}
       }
    \end{subfigure}
         \begin{subfigure}{0.31\textwidth}
        \centering
         
        \includegraphics[width=\textwidth]{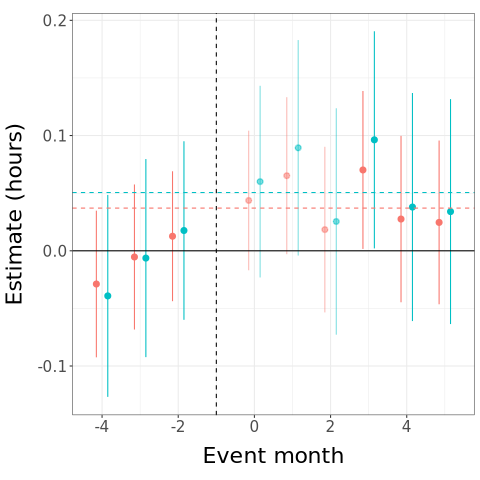}
       \caption{Word OOH Session Time
       % \label{fig:event_meetings}
       }
    \end{subfigure}
    \begin{subfigure}{0.4\textwidth}
        \centering
         
        \includegraphics[width=0.5\textwidth]{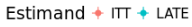}
       %\caption{Others' Doc. Read/Edited
       % \label{fig:event_meetings}
       %}
    \end{subfigure}\\

    \caption{Event Study Plots for the Effect of Copilot on Table \ref{table:app_results} Outcomes\label{fig:event_studies_appendix2}}
    \begin{tablenotes}
        These plots show the coefficients from regressions similar to Equation \eqref{regression} for the ITT (intent to treat) estimates and Equation \eqref{eqIV} for the LATE estimates. Vertical lines show 95\% confidence intervals based on worker-clustered standard errors. Horizontal dashed lines give the average estimated effect from Table \ref{table:app_results}. We include the coefficients for event months 0-2 for context.
    \end{tablenotes}
\end{figure}

\end{document}